\documentclass[sigconf,nonacm]{acmart}

\usepackage{colortbl}
\usepackage{xspace}
\usepackage[ruled, vlined, linesnumbered, longend]{algorithm2e}

\usepackage{utfsym}
\usepackage{paralist}
\usepackage{tikz}
\usepackage{framed}
\usetikzlibrary{fit,positioning,calc,math}

\newcommand{\gitlabrepo}{\url{https://version.helsinki.fi/ads/evaluating-learned-spatial-indexes/}}

\newcommand{\flood}{\textsc{Flood}\xspace}
\newcommand{\grid}{\textsc{GridFile}\xspace}
\newcommand{\kd}{\textsc{KDtree}\xspace}
\newcommand{\qd}{\textsc{QDtree}\xspace}
\newcommand{\rsmi}{\textsc{Rsmi}\xspace}
\newcommand{\str}{\textsc{STR}\xspace}
\newcommand{\rstar}{\textsc{R\text{*}tree}\xspace}
\newcommand{\cur}{\textsc{CUR}\xspace}
\newcommand{\rw}{\textsc{RWtree}\xspace}
\newcommand{\zindex}{\textsc{Zindex}\xspace}
\newcommand{\zm}{\textsc{ZMindex}\xspace}
\newcommand{\wazi}{\textsc{WAZI}\xspace}

\newcommand{\qdistr}{\ensuremath{\mathcal{Q}}\xspace}
\newcommand{\ddistr}{\ensuremath{\mathcal{D}}\xspace}
\newcommand{\uniformdist}{\textsc{Uniform}\xspace}
\newcommand{\clustercenters}{\ensuremath{ClusterMeans}\xspace}
\newcommand{\clustercentersshort}{\ensuremath{\mathcal{M}_D}\xspace}
\newcommand{\clustercovs}{\ensuremath{ClusterCovariances}\xspace}
\newcommand{\clustercovsshort}{\ensuremath{\mathcal{C}_D}\xspace}
\newcommand{\scalefactor}{\ensuremath{\kappa}\xspace}
\newcommand{\numclusters}{\ensuremath{C}\xspace}
\newcommand{\queryclustercenters}{\ensuremath{QueryClusterMeans}\xspace}
\newcommand{\queryclustercentersshort}{\ensuremath{\mathcal{M}_Q}\xspace}
\newcommand{\queryclustercovs}{\ensuremath{QueryClusterCovariances}\xspace}
\newcommand{\queryclustercovsshort}{\ensuremath{\mathcal{C}_Q}\xspace}

\AtBeginDocument{%
  }

\setcopyright{none}

\begin{document}

\title{Evaluating Learned Spatial Indexes}

\author{Sachith Pai}
\orcid{0009-0004-5128-7424}
\affiliation{%
  \institution{University of Helsinki}
  \city{Helsinki}
  \country{Finland}
}
\email{sachith.pai@helsinki.fi}

\author{Jun Yang}
\orcid{0009-0007-0187-0816}
\affiliation{%
  \institution{University of Helsinki}
  \city{Helsinki}
  \country{Finland}
}
\email{jun.yang@helsinki.fi}

\author{Michael Mathioudakis}
\orcid{0000-0003-0074-3966}
\affiliation{%
  \institution{University of Helsinki}
  \city{Helsinki}
  \country{Finland}
}
\email{michael.mathioudakis@helsinki.fi}

\renewcommand{\shortauthors}{Pai, Yang, and Mathioudakis}

\begin{abstract}
Learned indexes improve query performance by adapting search structures to data and workload distributions.
Although many learned indexes have been proposed, their trade-offs remain insufficiently understood for spatial range queries, where performance depends not only on model accuracy but also on data and query skew, layout granularity, selectivity, and storage behavior.
In this work, we perform an experimental study of learned indexes for spatial range queries.
We examine a representative set of indexes and address seven fundamental questions:
\begin{inparaenum}
  \item How does block size influence query latency, and what configurations yield optimal performance under varying selectivities?
  \item How do skewed data and query distributions impact index performance?
  \item How do indexes balance refinement and scan costs, and which designs favor one over the other?
  \item How do disk-based storage conditions alter optimal block size and latency trade-offs compared to in-memory settings?
  \item What are the construction costs of different indexes, and under what query volumes are these costs amortized?
  \item For a given data and query workload, which index is expected to perform best?
  \item Do index-selection insights learned from synthetic data generalize to real-world data distributions?
\end{inparaenum}
To enable the analysis, we use a framework with a common storage backend, standardized query execution pipelines, and controlled variations in data and query skew.

Our experiments reveal critical insights into refinement vs. scan trade-offs, the impact of block size, and the interplay between selectivity and layout effectiveness.
We synthesize these findings into a workload-based decision tree for index selection and validate it on real OpenStreetMap point sets with synthetic queries, confirming that its recommendations exhibit minimal decision regret and typically yield near-optimal query performance.

\end{abstract}

\maketitle

\section{Introduction}
Indexes are essential in database management systems (DBMSs), enabling efficient data retrieval for specific queries.
%
Because of their central role in performance, indexing methods are extensively studied, and new approaches continue to emerge.
%
Recently, advances in machine learning (ML) have inspired a new class of data-aware indexing techniques.
%
%
%
%
These methods, known as learned indexes \cite{Kraska_2018_CaseLearnedIndexa}, use compact predictive models to approximate data distributions or query costs, thereby tailoring the index structure to the underlying dataset.
%
Over the past few years, learned indexes have been proposed for diverse tasks, including one-dimensional search, spatial indexing, and nearest-neighbor queries.
Despite growing interest in learned indexes, comprehensive and systematic benchmarking remains limited~\cite{Kipf_2019_SOSDBenchmarkLearned,Liu_2025_HowGoodAre}.
Most existing studies introduce new learned indexing techniques and demonstrate improvements over selected baselines in narrowly defined scenarios.
However, the performance of learned indexes can vary significantly depending on factors such as data distribution, query workload, index granularity, and storage medium.
To draw reliable conclusions, we need careful comparisons under controlled conditions that reveal when learned indexes truly outperform traditional ones.
Such evaluations are essential for guiding future designs and enabling practical adoption.
While benchmarking frameworks exist for one-dimensional indexes, equivalent efforts for two-dimensional spatial indexing remain largely absent.
%

%
Spatial range queries introduce additional sources of variation because an index must both identify relevant spatial regions and control the amount of irrelevant data scanned within them, especially when data and query distributions are skewed.
As a result, comparisons depend not only on model accuracy, but also on data skew, query skew, layout granularity, query selectivity, and storage behavior.
This study systematically benchmarks learned spatial indexes for static two-dimensional point data under both in-memory and disk-backed range-query execution.
We evaluate a curated set of learned and traditional indexes within a unified benchmarking framework that uses a common storage backend, controlled synthetic workloads, and real OpenStreetMap validation data.
We focus on range queries, as they are the core operation for spatial indexes.
Our central thesis is that a rigorous comparative framework can reveal how learned and traditional indexes perform under different data and workload conditions for spatial range queries.
To make this comparison actionable, we also evaluate whether empirical findings can be translated into low-regret index choices, where a recommendation remains useful even if it does not always identify the single fastest index.
Our main contributions are:

\begin{enumerate}
    \item We analyze how block size influences performance and evaluate the impact of data and query skew on index performance.
    \item We analyze query latency by decomposing it into the refinement phase (identifying blocks that may contain matching points) and the scan phase (checking points within the identified blocks).
    \item We examine index performance with disk-based storage.
    \item We study when index construction costs are offset by improved query performance.
    \item We synthesize findings into a decision tree that guides index selection based on data and query workload characteristics.
    \item We validate the index-selection guidance on real-world OpenStreetMap point sets paired with synthetic query workloads.
\end{enumerate}

\section{Related Work}
Initial research on learned indexes concentrated primarily on the one-dimensional (1D) indexing problem.
As several learned methods emerged, each using distinct modeling approaches, there was a growing need for systematic evaluation.
This led to the development of multiple benchmark studies.
The SOSD benchmark~\cite{Kipf_2019_SOSDBenchmarkLearned} provided a foundational evaluation framework for learned indexes, comparing a range of indexes across diverse synthetic and real-world datasets.
Later studies expanded this scope by introducing comprehensive experimental evaluations that considered a wide range of learned indexes, data distributions, query workloads, and index update mechanisms \cite{Sun__LearnedIndexComprehensive,Marcus_2020_BenchmarkingLearnedIndexesa}.
Other works conducted targeted analyses of specific index families, particularly recursive model indexes, revealing performance bottlenecks in scalability and update efficiency~\cite{Maltry_2022_CriticalAnalysisRecursive}.
Several evaluations examined the readiness of updatable learned indexes, identifying limitations in their ability to handle dynamic workloads efficiently~\cite{Wongkham_2022_AreUpdatableLearneda, Ge_2023_CuttingLearnedIndex}.
Follow-up investigations explored the integration of updatable indexes into disk-resident database systems, revealing critical design trade-offs in in-memory layout and I/O behavior~\cite{Lan_2023_UpdatableLearnedIndexesa}.
Finally, recent work has examined the feasibility of efficiently constructing learned indexes, proposing sampling-based techniques to reduce training cost while preserving accuracy~\cite{Choi_2024_CanLearnedIndexes}.
These studies have collectively advanced the understanding of 1D learned indexes and enabled detailed comparisons across diverse operational settings.

However, most of these evaluations are limited to the 1D setting, with relatively little attention given to learned indexing in higher-dimensional or spatial contexts.
To our knowledge, there has been only one experimental survey, conducted by Liu et al.~\cite{Liu_2025_HowGoodAre}, which analyzed a collection of ten multidimensional learned indexes (including \zm, \rsmi, \flood, and \textsc{LISA}~\cite{DBLP:conf/sigmod/Li0ZY020}) on point-query workloads.
Their study finds that learned indexes often outperform classical indexes for in-memory range queries on points: for example, \flood and \textsc{LISA} were faster than R-tree variants across a variety of query selectivities, achieving notable speedups on large point sets.
Learned indexes also tend to use space more efficiently, yielding smaller memory footprints than R-tree indexes by storing model parameters instead of large node hierarchies.
However, the results also revealed some limitations.
Some methods, such as \rsmi, incur prohibitively high training costs at large scales, raising concerns about scalability.
Furthermore, for query types such as k-nearest neighbor (k-NN) search, learned indexes often fail to consistently outperform well-optimized traditional methods, such as \kd.
This mainly stems from the iterative probing required for k-NN queries, which reduces the benefit of learned predictions.
Overall, while learned indexes are well-suited for batched range queries on static point data, their performance advantage is not universal.

\begin{figure}[t]
  \captionsetup{skip=0pt}
  \centering
  \begin{tikzpicture}
    \tikzset{every node/.style={minimum size=1.2cm, inner sep=0pt}}
    \tikzset{mystyle/.style={line width=2pt,rounded corners,color=black!30!gray}}
    \tikzset{groupstyle/.style={thick, rounded corners=5pt,color=black!30!gray}}
    \tikzset{citenode/.style={line width=2pt,minimum height=12pt,minimum width=45pt}}

      \pgfmathsetmacro{\withingroup}{28.0}
      \pgfmathsetmacro{\acrossgroup}{34.0}

      \pgfmathsetmacro{\gridangle}{295}
      \pgfmathsetmacro{\floodangle}{\gridangle+\withingroup}

      \pgfmathsetmacro{\kdangle}{\floodangle+\acrossgroup-5}
      \pgfmathsetmacro{\qdangle}{\kdangle+\withingroup}

      \pgfmathsetmacro{\strangle}{\qdangle+\acrossgroup}
      \pgfmathsetmacro{\rwangle}{\strangle+\withingroup}
      \pgfmathsetmacro{\curangle}{\rwangle+\withingroup}
      \pgfmathsetmacro{\rsmiangle}{\curangle+\withingroup}
      \pgfmathsetmacro{\rstarangle}{\rsmiangle+\withingroup}

      \pgfmathsetmacro{\zangle}{\rstarangle+\acrossgroup-5}
      \pgfmathsetmacro{\waziangle}{\zangle+\withingroup-2}
      \pgfmathsetmacro{\zmangle}{\waziangle+\withingroup-2}


      \pgfmathsetmacro{\textdist}{2.5}
      \pgfmathsetmacro{\arcmountdist}{2}
      \pgfmathsetmacro{\segmentinner}{1.9}
      \pgfmathsetmacro{\segmentouter}{3.1}

      \draw
        (\gridangle:\textdist)node (GRID){\grid}
        (\floodangle:\textdist) node (FLOOD){\bf{\flood}}

        (\kdangle:\textdist) node (KD){\kd}
        (\qdangle:\textdist)node (QD){\bf{\qd}}

        (\rwangle:\textdist) node (RW){\bf{\rw}}
        (\strangle:\textdist) node (STR){\str}
        (\curangle:\textdist) node (CUR){\cur}
        (\rsmiangle:\textdist) node (RSMI){\bf{\rsmi}}
        (\rstarangle:\textdist) node (RSTAR){\rstar}

        (\zangle:\textdist) node  (Z){\zindex}
        (\waziangle:\textdist) node  (WAZI){\bf{\wazi}}
        (\zmangle:\textdist) node  (ZM){\bf{\zm}};

        \definecolor{floodcolor}{RGB}{34, 17, 80};
        \definecolor{wazicolor}{RGB}{152, 45, 128};
        \definecolor{rsmicolor}{RGB}{248, 118, 92};
        \definecolor{rwcolor}{RGB}{0, 0, 0};
        \definecolor{zmcolor}{RGB}{33, 145, 140};

      \begin{scope}[-]
        \draw[line width=0.6, solid, double, wazicolor] (\waziangle:\arcmountdist) .. controls ($(0,0)!0.5!($(\waziangle:\arcmountdist)!0.5!(\strangle:\arcmountdist)$)$) .. (\strangle:\arcmountdist);
        \draw[line width=0.6, solid, double, wazicolor] (\waziangle:\arcmountdist) .. controls ($(0,0)!0.5!($(\waziangle:\arcmountdist)!0.5!(\floodangle:\arcmountdist)$)$) .. (\floodangle:\arcmountdist);
        \draw[line width=0.6, solid, double, wazicolor] (\waziangle:\arcmountdist) .. controls ($(0,0)!0.5!($(\waziangle:\arcmountdist)!0.5!(\curangle:\arcmountdist)$)$) .. (\curangle:\arcmountdist);
        \draw[line width=0.6, solid, double, wazicolor] (\waziangle:\arcmountdist) .. controls ($(0,0)!0.75!($(\waziangle:\arcmountdist)!0.5!(\zangle:\arcmountdist)$)$) .. (\zangle:\arcmountdist);

        \draw[line width=1pt,solid,floodcolor] (\floodangle:\arcmountdist) .. controls ($(0,0)!0.5!($(\floodangle:\arcmountdist)!0.5!(\rstarangle:\arcmountdist)$)$) .. (\rstarangle:\arcmountdist);
        \draw[line width=1pt,solid,floodcolor] (\floodangle:\arcmountdist) .. controls ($(0,0)!0.75!($(\floodangle:\arcmountdist)!0.5!(\gridangle:\arcmountdist)$)$) .. (\gridangle:\arcmountdist);
        \draw[line width=1pt,solid,floodcolor] (\floodangle:\arcmountdist) .. controls ($(0,0)!0.75!($(\floodangle:\arcmountdist)!0.5!(\kdangle:\arcmountdist)$)$) .. (\kdangle:\arcmountdist);
        \draw[line width=1pt,solid,floodcolor] (\floodangle:\arcmountdist) .. controls ($(0,0)!0.5!($(\floodangle:\arcmountdist)!0.5!(\zangle:\arcmountdist)$)$) .. (\zangle:\arcmountdist);

        \draw[line width=1.8pt, loosely dashed, rsmicolor] (\rsmiangle:\arcmountdist) .. controls ($(0,0)!0.5!($(\rsmiangle:\arcmountdist)!0.5!(\gridangle:\arcmountdist)$)$) .. (\gridangle:\arcmountdist);
        \draw[line width=1.8pt, loosely dashed, rsmicolor] (\rsmiangle:\arcmountdist) .. controls ($(0,0)!0.5!($(\rsmiangle:\arcmountdist)!0.5!(\zmangle:\arcmountdist)$)$) .. (\zmangle:\arcmountdist);
        \draw[line width=1.8pt, loosely dashed, rsmicolor] (\rsmiangle:\arcmountdist) .. controls ($(0,0)!0.75!($(\rsmiangle:\arcmountdist)!0.5!(\rstarangle:\arcmountdist)$)$) .. (\rstarangle:\arcmountdist);
        \draw[line width=1.8pt, loosely dashed, rsmicolor] (\rsmiangle:\arcmountdist) .. controls ($(0,0)!0.5!($(\rsmiangle:\arcmountdist)!0.5!(\kdangle:\arcmountdist)$)$) .. (\kdangle:\arcmountdist);

        \draw[line width=1.5, densely dotted, zmcolor] (\zmangle:\arcmountdist) .. controls ($(0,0)!0.7!($(\zmangle:\arcmountdist)!0.5!(\rstarangle:\arcmountdist)$)$) .. (\rstarangle:\arcmountdist);

        \draw[line width=0.7pt, densely dashed, double, rwcolor ] (\rwangle:\arcmountdist) .. controls ($(0,0)!0.5!($(\rwangle:\arcmountdist)!0.5!(\rstarangle:\arcmountdist)$)$) .. (\rstarangle:\arcmountdist);

      \end{scope}

\draw[groupstyle] (277.0:\segmentinner) -- (285.0:\segmentinner) -- (293.0:\segmentinner) -- (301.0:\segmentinner) -- (309.0:\segmentinner) -- (317.0:\segmentinner) -- (325.0:\segmentinner) -- (333.0:\segmentinner) -- (333.0:\segmentouter) -- (325.0:\segmentouter) -- (317.0:\segmentouter) -- (309.0:\segmentouter) -- (301.0:\segmentouter) -- (293.0:\segmentouter) -- (285.0:\segmentouter) -- (277.0:\segmentouter) --  cycle;
\draw[groupstyle] (344.0:\segmentinner) -- (352.0:\segmentinner) -- (360.0:\segmentinner) -- (368.0:\segmentinner) -- (376.0:\segmentinner) -- (384.0:\segmentinner) -- (392.0:\segmentinner) -- (392.0:\segmentouter) -- (384.0:\segmentouter) -- (376.0:\segmentouter) -- (368.0:\segmentouter) -- (360.0:\segmentouter) -- (352.0:\segmentouter) -- (344.0:\segmentouter) --  cycle;
\draw[groupstyle] (406.0:\segmentinner) -- (414.0:\segmentinner) -- (422.0:\segmentinner) -- (430.0:\segmentinner) -- (438.0:\segmentinner) -- (446.0:\segmentinner) -- (454.0:\segmentinner) -- (462.0:\segmentinner) -- (470.0:\segmentinner) -- (478.0:\segmentinner) -- (486.0:\segmentinner) -- (494.0:\segmentinner) -- (502.0:\segmentinner) -- (510.0:\segmentinner) -- (518.0:\segmentinner) -- (526.0:\segmentinner) -- (534.0:\segmentinner) -- (534.0:\segmentouter) -- (526.0:\segmentouter) -- (518.0:\segmentouter) -- (510.0:\segmentouter) -- (502.0:\segmentouter) -- (494.0:\segmentouter) -- (486.0:\segmentouter) -- (478.0:\segmentouter) -- (470.0:\segmentouter) -- (462.0:\segmentouter) -- (454.0:\segmentouter) -- (446.0:\segmentouter) -- (438.0:\segmentouter) -- (430.0:\segmentouter) -- (422.0:\segmentouter) -- (414.0:\segmentouter) -- (406.0:\segmentouter) --  cycle;
\draw[groupstyle] (547.0:\segmentinner) -- (555.0:\segmentinner) -- (563.0:\segmentinner) -- (571.0:\segmentinner) -- (579.0:\segmentinner) -- (587.0:\segmentinner) -- (595.0:\segmentinner) -- (603.0:\segmentinner) -- (611.0:\segmentinner) -- (619.0:\segmentinner) -- (627.0:\segmentinner) -- (627.0:\segmentouter) -- (619.0:\segmentouter) -- (611.0:\segmentouter) -- (603.0:\segmentouter) -- (595.0:\segmentouter) -- (587.0:\segmentouter) -- (579.0:\segmentouter) -- (571.0:\segmentouter) -- (563.0:\segmentouter) -- (555.0:\segmentouter) -- (547.0:\segmentouter) --  cycle;


  \node [matrix,very thick,minimum height=20pt] (my matrix) at (3.6,0)
  {
    \node[citenode] {\cite{DBLP:conf/sigmod/NathanDAK20}}; \\ \draw[line width=1pt,solid,floodcolor] (-0.45,0) -- (0.45,0) {}; \\ [6pt]
    \node[citenode] {\cite{DBLP:conf/edbt/PaiM024}}; \\ \draw[line width=0.6, solid, double, wazicolor]   (-0.45,0) -- (0.45,0) {}; \\ [6pt]
    \node[citenode] {\cite{Qi_2020_EffectivelyLearningSpatiala}}; \\ \draw[line width=1.8pt, loosely dashed, rsmicolor] (-0.45,0) -- (0.45,0) {}; \\ [6pt]
    \node[citenode] {\cite{Wang_2019_LearnedIndexSpatiala}}; \\ \draw[line width=1.5, densely dotted, zmcolor] (-0.45,0) -- (0.45,0) {}; \\ [6pt]
    \node[citenode] {\cite{Dong_2022_RWTreeLearnedWorkloadawarea}};\\  \draw[line width=0.7pt, densely dashed, double, rwcolor] (-0.45,0) -- (0.45,0) {}; \\
  };

  \end{tikzpicture}
  \caption{Chord diagram showing the indexes evaluated in this work (nodes in the diagram) and prior comparisons between them (edges). Learned indexes are shown in bold. Our work aims to fill the missing comparisons in the analysis of learned spatial indexes.}
  \Description{A chord diagram with evaluated index names arranged around a circle and grouped by index family. Curved edges connect index pairs that were compared in prior work, with edge styles keyed to the cited studies and learned index names shown in bold.}
  \label{fig:chord_diagram}
  \vspace{-3pt}
\end{figure}
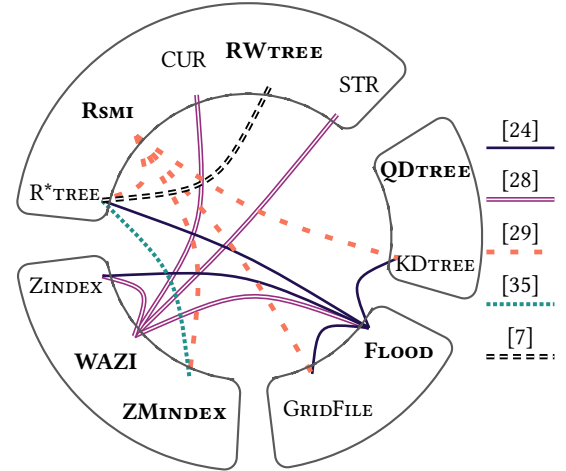

Figure~\ref{fig:chord_diagram} summarizes this gap by showing which index pairs have been compared in prior learned-spatial-index studies and which comparisons remain missing.
While our work is similar in scope to Liu et al.~\cite{Liu_2025_HowGoodAre}, it differs in five crucial ways that address previously unexamined benchmarking shortcomings in spatial learned indexing:
\begin{inparaenum}
    \item We re-implement the methods evaluated in our experiments to avoid implementation-induced biases, whereas their analysis used original implementations and open-source libraries.
    \item We analyze all methods within a unified framework, which stores data points in blocks with a common structure across indexes.
    \item We construct diverse synthetic datasets and query workloads to systematically evaluate index behavior under varying levels of skew and selectivity, an aspect not addressed in previous work.
    \item We present detailed analyses of performance-affecting factors not covered in previous studies, including block size (Section~\ref{subsec:optimal_block_sizes}), the breakdown of latency into refinement and scan phases (Section~\ref{subsec:refinement_vs_scan.}), effects of data and query distributions (Section~\ref{subsec:effect_data_query_skew}), and performance on disk-based data (Section~\ref{subsec:disk_based_data}).
    \item We synthesize our empirical findings into a decision tree (Figure~\ref{fig:decision_tree}) that provides practical guidance for selecting the most suitable index under given workload characteristics.
\end{inparaenum}
Together, these differences allow this work to address important questions about learned spatial indexes that have not been previously examined.

\section{Compared Indexes} \label{sec:baselines}

We evaluate both traditional and learned indexes, grouped by design principle into Grid-based Partitioning Indexes, Space Partitioning Indexes, Data Partitioning Indexes, and Ordering-based Indexes.
The following sections describe each method, its implementation choices, and any learned components used.
Table~\ref{tab:baselines} summarizes the key characteristics of each index.

\begin{table}[t!]
  \captionsetup{skip=-15pt}

  \setlength{\tabcolsep}{1pt}
    \centering
    \begin{tabular}{ccc}
      \toprule
      Index & Query-Aware & Learned Component  \\
      \midrule
      \grid~\cite{DBLP:conf/eci/NievergeltHS81}     &   No & - \\
      \bf{\flood}~\cite{DBLP:conf/sigmod/NathanDAK20}    &   Yes & Surrogate latency predictor. \\
      \hline
      \kd~\cite{Bentley_1975_MultidimensionalBinarySearch}    &   No & -  \\
      \bf{\qd}~\cite{Yang_2020_QdtreeLearningDataa}   &   Yes & RL-agent for data splitting. \\
      \hline
      \str~\cite{Leutenegger_1997_STRSimpleEfficienta}    &   No & -  \\
      \rstar~\cite{DBLP:conf/sigmod/BeckmannKSS90}    &   No & -  \\
      \bf{\rsmi}~\cite{Qi_2020_EffectivelyLearningSpatiala}     &   No & Neural network at each node. \\
      \cur~\cite{Ross_2001_CostbasedUnbalancedRtrees}     &   Yes & -  \\
      \bf{\rw}~\cite{Dong_2022_RWTreeLearnedWorkloadawarea}    &   Yes & Learned 4-D MBR-overlap \\ & & cost estimator.  \\
      \hline
      \zindex~\cite{tropf1981multidimensional,DBLP:journals/gidr/Markl00,DBLP:conf/wwca/Bayer97}    & No & -  \\
      \bf{\zm}~\cite{Wang_2019_LearnedIndexSpatiala}    & No & Learned 1-D index (RMI). \\
      \bf{\wazi}~\cite{DBLP:conf/edbt/PaiM024}  & Yes & Data/query density estimator.  \\
      \bottomrule
    \end{tabular}
    \caption{List of indexes compared, grouped into categories, with the learned indexes shown in bold.}
    \Description{A table grouping the evaluated indexes into space-partitioning, data-partitioning, order-based, and sample-based categories, with learned indexes highlighted in bold.}
    \label{tab:baselines}
  \end{table}

\subsection{Grid-based Partitioning Indexes}
Grid-based Partitioning Indexes divide the data space into a regular grid of cells.
Each point is assigned directly to the cell containing it.

\subsubsection{\grid~\cite{DBLP:conf/eci/NievergeltHS81}}
The \grid partitions the search space with a fixed grid layout.
Points are grouped into blocks by their grid cells, and the number of splits per dimension is chosen so that blocks contain the required average number of points.

\subsubsection{\flood~\cite{DBLP:conf/sigmod/NathanDAK20}}
The \flood index uses an optimized grid layout that minimizes query latency for a given data distribution and query workload.
It applies surrogate-based optimization~\cite{Neufang_2024_SurrogateBasedOptimizationTechniques,Queipo_2005_SurrogatebasedAnalysisOptimization}, training a surrogate ML model on measured query latencies for a small set of grid layouts to predict the latency of unseen layouts.
These predictions serve as the objective for evaluating candidate layouts.

\subsubsection{Implementation}
The Grid-based Partitioning Indexes are implemented as two-level search trees over vertical and horizontal partitioning boundaries.
For \grid, we choose equal vertical and horizontal splits independently to form rows and columns.
Our \flood implementation uses the same structure, but directly evaluates candidate layouts on a data sample instead of training a surrogate model.
We avoid surrogate optimization because, in our two-dimensional in-memory setting, surrogate training costs more than simulating candidate query latencies.

\subsection{Space Partitioning Indexes}

Space Partitioning Indexes recursively divide the data space into smaller subregions.
Each node represents a domain region, producing a hierarchical spatial decomposition for index construction.

\subsubsection{\kd~\cite{Bentley_1975_MultidimensionalBinarySearch}}
The \kd recursively splits the data space along alternating dimensions using median splits, so the points in the current region are divided into two roughly equal subsets.
This process continues until each cell has at most the target block size, producing a binary tree.

\subsubsection{\qd~\cite{Yang_2020_QdtreeLearningDataa}}
The \qd index customizes \kd by learning the split dimension and location to minimize block accesses.
Its training algorithm uses a reinforcement learning agent to select these decisions at each node while constructing candidate trees.
The completed tree is rewarded according to block-skipping efficiency, allowing the agent to learn better partitioning policies over time.

\subsubsection{Implementation}
The Space Partitioning Indexes are implemented as binary trees, with each node defined by a split dimension and split location.
In the case of \kd, the split dimension alternates at each level of the tree, and the split location is selected as the median value along the chosen dimension.
\qd is implemented as an optimized \kd, using a sampling-based strategy to select both the split dimension and location while minimizing the specified objective function. %
We also experimented with the RL construction method from \cite{Yang_2020_QdtreeLearningDataa}, which learns a split policy from a finite set of candidate split locations.
In our implementation, the agent frequently appeared to settle in local optima, yielding policies close to fixed split rules rather than reliably data-dependent decisions.
In our setting, the RL approach was also feasible only for shallow trees, since deeper trees require many candidate split locations and make training prohibitively expensive.
We therefore use sampling-based optimization for \qd.

\subsection{Data Partitioning Indexes}
Data Partitioning Indexes group points into subsets represented by minimum bounding rectangles (MBRs) arranged hierarchically.
They reduce query cost by optimizing point partitions and MBR overlap.
Classic R-trees~\cite{DBLP:conf/sigmod/Guttman84} build these structures iteratively, inserting objects into leaf nodes while minimizing the enlargement of MBRs along the path.
Below, we describe several variants and learned extensions.

\subsubsection{\str~\cite{Leutenegger_1997_STRSimpleEfficienta}}
The Sort-Tile Recursive (\str) algorithm bulk-loads R-trees by sorting points along one dimension, partitioning them into equal-sized tiles, and repeating along alternating dimensions until blocks reach the target size.
This produces compact trees with low MBR overlap.

\subsubsection{\rstar~\cite{DBLP:conf/sigmod/BeckmannKSS90}}
The \rstar refines R-tree insertion by considering perimeter and perimeter change in addition to area enlargement \cite{Beckmann_2009_RevisedRtreeComparison}.
This favors square-like MBRs, reducing node overlap and query cost.

\subsubsection{\rsmi~\cite{Qi_2020_EffectivelyLearningSpatiala}}
The Recursive Spatial Model Index (\rsmi) is a learned R-tree-style index.
It recursively partitions spatial data into regions handled by learned models rather than explicit bounding boxes.
\rsmi first maps data into rank space, where each dimension is replaced by rank order, then applies a space-filling curve to sort and group points into blocks.
At each stage, a learned function routes points to sub-models, mirroring R-tree splits with neural networks.
Queries traverse this model hierarchy to identify leaf blocks for scanning.

\subsubsection{\cur~\cite{Ross_2001_CostbasedUnbalancedRtrees}}
The Cost-based Unbalanced R-tree (\cur-tree) relaxes the R-tree's strict balancing constraints so the tree shape can reflect expected node access frequency.
\cur uses a workload-derived cost heuristic for data points.
During insertion, it evaluates local reorganizations, including splits, promotions, and demotions, and keeps changes that lower expected query cost.
This adapts the tree to workload patterns.
This often improves performance in in-memory settings where computation and cache behavior dominate.

\subsubsection{\rw~\cite{Dong_2022_RWTreeLearnedWorkloadawarea}}
The \rw extends \rstar with workload awareness.
It augments top-down insertion and splitting with a learned cost model that estimates the trade-off of placing new entries in each child node.
During insertion, it predicts the subtree or split strategy that best reduces future query cost, often favoring layouts aligned with high-demand regions.
This reduces MBR overlap in frequently queried regions and lowers query cost.

\subsubsection{Implementation}
Our implementation of \rstar uses the node selection mechanism described in~\cite{DBLP:conf/sigmod/BeckmannKSS90}.
For \rw, we implement a simplified learned cost model following~\cite{Dong_2022_RWTreeLearnedWorkloadawarea}.
We encode each candidate MBR, either an existing node MBR or an expanded-node MBR, as a four-dimensional point using its lower and upper bounds, and use a learned density estimator \cite{Wen_2022_RandomForestDensitya} to estimate query-overlap cost.
During insertion, \rw selects the subtree whose expanded MBR yields the smallest predicted overlap-cost increase for the new point.
\str~\cite{Leutenegger_1997_STRSimpleEfficienta} partitions the data into equal-sized subsets by alternately splitting along each dimension at each level of the tree.
\cur~\cite{Ross_2001_CostbasedUnbalancedRtrees} extends \str by replacing equal-sized splits with query-weighted splits, where partitions contain an equal estimated query cost rather than an equal number of points.
We compute query-weighted splits by weighting each point by the number of queries that access it and binary-searching cumulative weights for split locations.
The \rsmi index is implemented in Python using PyTorch, following the construction procedure outlined in~\cite{Qi_2020_EffectivelyLearningSpatiala}.
After construction, we store the learned data partitions (the bounding boxes of each partition) and the data points within each leaf-level node.

\subsection{Ordering-Based Indexes}
Ordering-Based Indexes linearize spatial coordinates into a one-dimensional order.
This makes standard one-dimensional indexes applicable while preserving spatial relationships through space-filling curves or learned mappings.

\subsubsection{\zindex~\cite{tropf1981multidimensional,DBLP:journals/gidr/Markl00,DBLP:conf/wwca/Bayer97}}
\zindex uses a Z-order space-filling curve to linearize points, then indexes the resulting one-dimensional values with a structure such as a B-tree.

\subsubsection{\zm~\cite{Wang_2019_LearnedIndexSpatiala}}
\zm extends \zindex by replacing the B-tree with a Recursive Model Index \cite{Kraska_2018_CaseLearnedIndexa}.
\zm combines the compactness of learned indexes with the ordering efficiency of space-filling curves.

\subsubsection{\wazi~\cite{DBLP:conf/edbt/PaiM024}}
\wazi learns a workload-aware space-filling curve based on Z-order.
\wazi uses learned density estimates and a Z-order cost function to optimize the layout for query processing.
It also exploits Z-order monotonicity to skip large data ranges irrelevant to a query.

\subsubsection{Implementation}
We use the publicly available implementations of \zindex and \wazi from \cite{DBLP:conf/edbt/PaiM024}.
For \zm~\cite{Wang_2019_LearnedIndexSpatiala}, we adapt the original design to our paged architecture: after sorting by Z-order, we group points into fixed-size blocks, store each block's minimum and maximum Z-order values, and index those ranges with a PGM-index~\cite{Ferragina_2020_PGMindexFullydynamicCompressed}, replacing the original RMI~\cite{Kraska_2018_CaseLearnedIndexa}.

\section{Scope of Our Study}

In this section, we outline the key questions that guide our experimental evaluation, the context in which this evaluation is performed, and the experimental parameters considered.
To set the stage, we first present the motivation behind our research questions and provide an overview of our experimental plan.
We then describe the synthetic data generation process used to construct datasets and query workloads for our index evaluation.
Finally, we provide a high-level summary of index implementation and experimental setup details, establishing a basis for the thematic sections that follow.

\subsection{Research Questions}

\subsubsection{Block Size}
We first examine the effect of block size, also known as leaf size, which is one of the most influential parameters in spatial indexing.
It controls the granularity of the data layout and dictates how many points reside in each leaf node.
A small block size increases the number of index nodes, requiring more query-processing steps to locate relevant blocks but reducing the number of irrelevant points examined.
Conversely, a large block size reduces the number of lookup steps but risks scanning many unnecessary points.
Despite its importance, most prior studies either fix block size arbitrarily or assume defaults inherited from database systems without questioning their suitability for different index families.
For learned indexes, where layout decisions are closely tied to modeling assumptions, block size may interact with model accuracy and workload sensitivity in unexpected ways.
Understanding these interactions is critical both for fair benchmarking and for guiding practical index tuning.

\noindent\textbf{RQ1:}\textit{
    How does block size influence query latency across indexes, and which configurations yield optimal performance under varying query selectivities?
}
We answer this research question by systematically varying block size across all index types, measuring query latencies, and identifying sweet-spot configurations.
This reveals how each index family responds to block-size tuning and whether learned indexes adapt better than traditional indexes.

\subsubsection{Effect of Data and Query Distribution}
The advantage of learned indexes lies in their ability to adapt to the underlying data distribution.
In particular, their ability to efficiently model skewed data distributions is key to improving performance, yet this aspect has not been evaluated systematically.
Prior evaluations often highlight strong performance on datasets such as OSM~\cite{OSM} and TPC-H~\cite{TPCH} but rarely isolate the specific role of skew in shaping outcomes.
Without controlled experiments, it remains unclear whether improvements come from modeling advantages or from other factors coinciding with dataset characteristics.

In addition to data skew, query skew introduces a different dimension of performance variability that is equally important to consider.
Even when data is uniformly distributed, skewed query patterns—where some regions are queried more frequently—can stress certain index regions disproportionately.
This is particularly relevant for query-agnostic learned indexes, which optimize only for the data distribution.
Analyzing index behavior under varying levels of query skew reveals robustness in handling workload variability, making query-skew analysis essential for evaluating indexes.

\noindent\textit{\textbf{RQ2:}
    How do skewed data and query distributions affect the performance of learned indexes?
    }

We answer this research question by generating synthetic datasets and workloads with precisely controlled levels of skew using Gaussian Mixture Models.
By separating the effects of data skew and query skew, we assess robustness and identify which indexes gracefully handle variability in data and query skew.

\subsubsection{Refinement Latency versus Scan Latency}
Every range query can be decomposed into two phases: refinement, where the index identifies potentially relevant blocks, and scanning, where those blocks are filtered to produce results.
The relative cost of these phases reveals how an index achieves efficiency.
Some structures minimize refinement effort by simplifying search, but at the cost of scanning many irrelevant points.
Others focus on aggressive pruning, potentially incurring high refinement overhead.
Surprisingly, most evaluations report only total latency, obscuring whether gains come from refinement, scanning, or both.
Without this breakdown, it is impossible to understand why one index outperforms another, or under what conditions design choices pay off.
\noindent\textit{\textbf{RQ3:}
How do indexes balance refinement and scan phases, and which designs optimize one over the other?
}

We answer this research question by separately measuring refinement and scan latencies for each index across configurations.
This breakdown reveals the two levels of optimization possible within learned indexes: modeling the data distribution to minimize refinement, and optimizing block layout to minimize scan.

\subsubsection{Indexing Disk-Based Data}
While in-memory experiments dominate the literature on learned indexes, many real-world systems operate under disk-based storage constraints~\cite{Abu-Libdeh_2020_LearnedIndexesGooglescaleDiskbased}.
Disk access reshapes performance dynamics: refinement remains relatively cheap, but scanning incurs orders of magnitude higher latency due to I/O bottlenecks.
These conditions can shift the optimal trade-offs observed in-memory, making previously efficient designs suboptimal.
Empirical evidence on learned indexes with disk-resident data is limited, with only a few published disk-based evaluations.
Understanding the disk-based performance of learned indexes is crucial for assessing their practical viability.

\noindent\textit{\textbf{RQ4:}
How do disk-based storage conditions alter the optimal block size and the refinement-scan latency trade-off compared to in-memory settings?
}

We answer this research question by repeating our experiments with data stored on disk via memory mapping and contrasting them with in-memory results.
This reveals how storage medium changes the landscape of optimal configurations and which indexes are resilient across both settings.

\subsubsection{Index Construction Cost and Redemption}
Query performance is only one aspect of index utility.
Construction cost is also critical, especially in systems where data changes frequently or indexes must be rebuilt often.
Learned indexes typically require expensive training and layout optimization, raising the question of whether their query-time benefits justify their upfront construction overhead.
In practice, a method that accelerates queries but takes days to build may be unsuitable except at very large scales.
Yet construction costs are often reported without systematic comparison or analysis of when they “pay for themselves.”
Filling this gap in the literature is essential for moving from theoretical performance to practical deployment.
\noindent\textit{\textbf{RQ5:}
    What are the construction costs of different indexes, and under what query volumes are these costs effectively amortized?
}

We answer this research question by measuring the build times and calculating the “redemption” threshold, i.e., the number of queries required to offset construction overhead.
This yields a practical perspective on which indexes suit short-lived tasks and which are viable only for massive workloads.

\subsubsection{Index Selection}
Ultimately, practitioners seek to answer a fundamental question: which index should they use for a specific dataset and workload?
Yet prior evaluations typically stop short of actionable guidance, instead presenting raw performance numbers.
The reality is that no single index dominates across all conditions—performance depends on skew, selectivity, and storage environment.
Without a systematic selection method, users face trial-and-error or rely on intuition.
A principled way of translating workload characteristics into index choice would make research insights directly usable in practice.

\begin{figure*}[t!]
    \captionsetup{skip=5pt}
    \centering
    {\includegraphics[clip,width=0.75\linewidth]{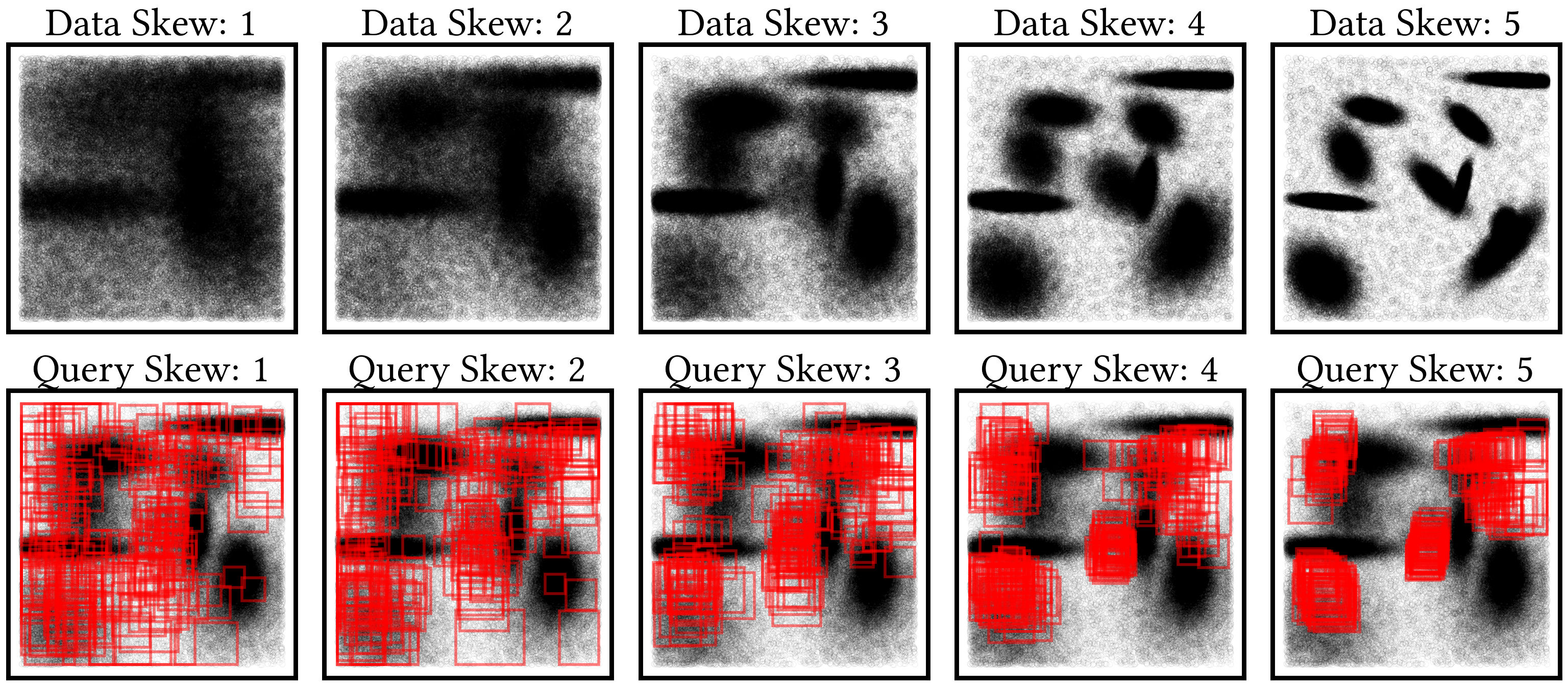}}
    \caption{
        Example datasets and query workloads generated from Gaussian mixture models (GMMs). Top: data points with increasing levels of skew drawn from the same underlying GMM.
Bottom: corresponding query distributions generated from a GMM aligned with the data points (shown here for data skew = 3).
        }
    \Description{A two-row panel of scatter plots. The top row shows synthetic point datasets becoming more clustered as skew increases, and the bottom row shows corresponding query workloads generated from aligned query distributions.}
    \label{fig:data_query_clustering_example}
    \vspace{-3pt}
\end{figure*}

\noindent\textit{\textbf{RQ6:}
 Given data and query workload statistics, which index is expected to provide the best range query performance?
 }

We address this question by ranking indexes across experimental configurations and distilling the observed patterns into a decision tree.
This offers a practical, data-driven guide to index selection, bridging the gap between experimental evaluation and deployment.

\subsubsection{Validation of Synthetic-Data Insights}
Synthetic benchmarks allow us to vary skew and selectivity in a controlled manner, but they also raise an important question of external validity.
The practical value of the insights derived from our synthetic analysis depends on whether they remain predictive when the underlying spatial points come from real-world data.
To test this, we conduct a validation experiment that combines real OpenStreetMap point sets with synthetic query workloads generated from Gaussian mixture models.
This hybrid setting preserves control over the query distribution while evaluating whether the index-selection guidance learned from synthetic data transfers to more realistic spatial layouts.

\noindent\textit{\textbf{RQ7:}
Do the index-selection insights learned from synthetic data generalize to real-world data distributions?
}

We answer this research question by evaluating the learned decision tree on 25 validation datasets, each containing 8 million real-world OpenStreetMap points and paired with 1,000 synthetic queries drawn from Gaussian mixture models.
This measures whether the selection rules extracted from controlled synthetic experiments remain useful when applied to real spatial point configurations.

\subsection{Dataset and Query Workloads}

\begin{algorithm}[t]
    \caption{Generate Data Points \ddistr and Query Workloads \qdistr}
    \Description{Pseudocode for sampling clustered data and query workloads by drawing Gaussian mixture components, scaling covariance matrices by skew factors, and returning the generated distributions.}
    \label{algo:generatedata}

    \SetKwInOut{inputargs}{Input}
    \SetKwInOut{result}{Output}
    \SetKwInOut{return}{Return}

    \SetKwFunction{sampl}{Sample}
    \SetKwFunction{scale}{Scale}
    \SetKwFunction{samplgaussians}{SampleMixGaussian}

    \SetKwFunction{SolveCell}{SolveCell}
    \SetKwFunction{Greedy}{Greedy}
    \SetKwFunction{uniformsearch}{UniformSearchCandidates}
    \SetKwFunction{split}{Split}

    \inputargs{number-of-clusters \numclusters, clusteredness scale}
    \result{A set of datasets with varying clusteredness \ddistr and a set of queries for each dataset \qdistr}

    \clustercenters\clustercentersshort:= \sampl(\uniformdist, \numclusters) \\
    \clustercovs\clustercovsshort:= Sample random 2D covariance matrices for each cluster. \\
    \ForEach{Clusteredness scale $\scalefactor_D$}{
        $\ddistr_i$ = \samplgaussians(\clustercentersshort, \scale( \clustercovsshort, $\scalefactor_D$))\\
        \tcp{Sample the query cluster centers from data}
        \queryclustercenters \queryclustercentersshort:= \sampl($\ddistr_i$, \numclusters) \\
        \queryclustercovs\queryclustercovsshort:= Sample random 2D covariance matrices for each cluster. \\
        \ForEach{Clusteredness scale $\scalefactor_Q$}{
            $\qdistr_{i,j}$ = \samplgaussians(\queryclustercentersshort, \scale(\queryclustercovsshort, $\scalefactor_Q$))\\
            $\qdistr_i$ += \{ $\qdistr_{i,j}$ \} \\
        }

        $\ddistr$ += \{ $\ddistr_i$ \} \\
        $\qdistr$ += \{ $\qdistr_i$ \} \\
    }
    \return{\ddistr , \qdistr}
\end{algorithm}

We generate synthetic datasets and query workloads using Gaussian Mixture Models (GMMs).
Our primary focus is on density-based skew, where data points or queries are unevenly distributed across the spatial domain.
To quantify this skew, we use a grid-based metric called Spatial Density Entropy (SDE) \cite{Liu_2018_AddingSpatialDistribution,Liu_2018_FastIdentificationUrban,Zhang_2022_CARMICacheawareLearneda}.
SDE is calculated by overlaying a uniform grid over the data space, counting the number of data points within each grid cell, and computing the entropy of this distribution.
The maximum SDE is achieved when all grid cells contain an equal number of points, corresponding to a uniform distribution, whereas low values indicate increasingly skewed data.
To construct a dataset, we begin by randomly sampling the number of Gaussian components in the GMM\@.
The mean of each component is sampled uniformly from the unit square, while the covariance matrices are sampled randomly to capture a range of spread patterns.
From each GMM configuration, we draw sample points and discard any that fall outside the unit square.
This procedure allows us to generate datasets with fine-grained control over the level and structure of clustering.
To construct query workloads, we follow a similar approach to dataset generation.
Specifically, we sample query centers from a GMM and expand each center into a rectangular query window to meet a target selectivity.
Unlike data generation, the means of the Gaussian components for the query GMM are not sampled uniformly.
Instead, they are sampled from the actual data points in the dataset, ensuring that the query distribution is correlated with the data distribution.
This design mimics real-world scenarios where query patterns often align with dense or semantically important regions in the dataset, such as shopping districts in a city map or popular tourist attractions in a geographical dataset.
The covariance matrices for the query GMMs are again sampled randomly to introduce variability in query shapes and orientations.
To simulate a range of skew levels, we scale the covariance matrices of the Gaussian components along their principal axes.
This scaling allows us to control the degree of clustering while preserving the overall structure of the GMM.
From each parameterization, we generate five datasets with different levels of data skew.
These datasets enable us to isolate the effect of skew while maintaining the same underlying GMM structure.
Figure~\ref{fig:data_query_clustering_example} (top) visualizes a set of such datasets generated from a shared GMM under varying skew levels.
Similarly, we generate five sets of query workloads for each dataset, varying only the clustering parameter.
Figure~\ref{fig:data_query_clustering_example} (bottom) shows one example of query distributions derived from the same underlying data.
This structured generation procedure allows us to independently control and analyze the effects of both data and query skew on index performance.
Table~\ref{tab:expsetup:dataset_params} summarizes the parameters used in our data and query generation process.
Algorithm~\ref{algo:generatedata} describes the process of generating both datasets and query workloads, showcasing how parameters are sampled and scaled to achieve the desired skew levels.

\subsection{Implementation and Experimental Setup}

The index structures evaluated in our experiments are implemented with a unified storage backend.
This backend stores data as a collection of blocks, which function similarly to the leaf nodes in traditional indexing schemes.
Range query processing in each index is executed in two distinct phases.
The \emph{refinement phase} involves leveraging the internal information maintained by the index to identify a relevant subset of blocks.
In the subsequent \emph{scan phase}, these selected blocks are accessed, and the contained data points are filtered to produce the final query result.

All index structures are implemented in C++ as search structures that, given a query, return a refined list of data blocks.
A common block-based storage layer is used for all indexes to store and retrieve data efficiently for scanning.
For experiments involving disk-based data, this block storage is backed by a memory-mapped file to simulate realistic access patterns while preserving a consistent interface and layout across index types.
Specifically, we construct the index in memory and store the blocks in a memory-mapped file at the end of index construction.
For indexes with learned components, we use a 10\% random sample of the dataset and associated query workload for training.
In the case of \rsmi, training is performed in Python using PyTorch.
After the model is trained, the resulting index structure—comprising learned partition boundaries and block assignments—is exported and used as a read-only search index in C++ for evaluation.

\begin{table}[t]
    \centering
    \captionsetup{skip=2pt}
    \setlength{\tabcolsep}{2pt}
    \caption{Experimental parameters; block size is optimized per index and configuration from the options listed here, and 0.1024\% is the default selectivity unless varied.}
    \Description{A table listing dataset size, query selectivity values, query workload size, average data and query SDE levels, and candidate block sizes used in the experiments.}
    \begin{tabular}{cc}
        \toprule
        Parameter &  Values (default in bold) \\
        \midrule
        Dataset size        &     8 $\times 10^6$ \\
        Query selectivity (\%) & [0.0064, 0.0256, \underline{0.1024}, 0.4096, 1.6384] \\
        Query workload size & 10,000 \\
        Avg. data SDE per level & [0.985 0.975 0.958 0.924 0.870] \\
        Avg. query SDE per level & [0.963 0.933 0.890 0.830 0.755] \\
        Block size & [32, 64, 128, 256, 512, 1024, 2048, 4096] \\
        \bottomrule
    \end{tabular}
    \label{tab:expsetup:dataset_params}
    \vspace{-3pt}
\end{table}

All experiments are conducted on a machine equipped with an Intel Xeon E5-2680v4 CPU @ 2.40GHz and 16 GB of RAM.
All evaluations are single-threaded, with no GPU acceleration or parallelism used, to ensure consistency in latency measurements and to reflect typical CPU-bound deployments.
We performed detailed experiments on a dataset of 8 million data points across block-size, query-selectivity, data-skew, and query-skew settings.
We use these detailed experimental results to answer the research questions introduced above.
Table~\ref{tab:expsetup:dataset_params} summarizes the complete set of parameters used in our experiments.

\section{Experimental Results}
\subsection{Optimal Block Size} \label{subsec:optimal_block_sizes}

\begin{figure*}[t]
  \centering
  \captionsetup{skip=2pt}
  {\includegraphics[trim={0 0.2cm 0 0.0cm},clip,width=0.9\textwidth]{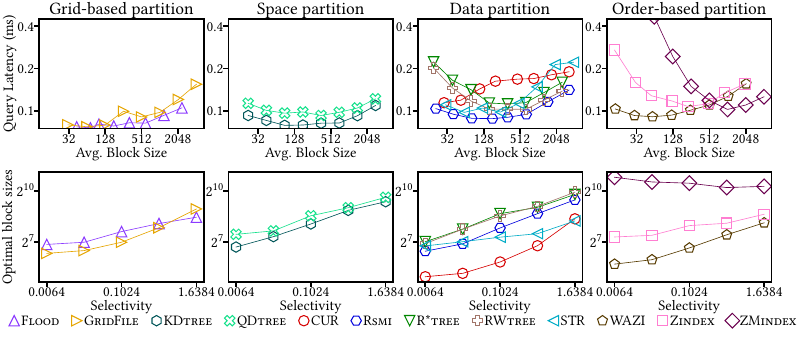}} %
        \caption{Query latency across block sizes (top) and optimal block size across selectivities (bottom). Realized average block size is about 73\% of configured block size because several construction algorithms split data non-uniformly. The optimal block sizes are identified by local minima (sweet-spot), marking tuned granularities.}
        \Description{A two-part plot comparing index behavior under block-size tuning. The upper panels show query latency curves across configured block sizes, and the lower panels show the selected optimal block size as query selectivity changes.}
  \label{fig:best_blocksize_plus_effect_of_selectivity}
  \vspace{-3pt}
\end{figure*}

\begin{figure}[t!]
\captionsetup{skip=5pt}
  \centering
  {\includegraphics[trim={0 0cm 0 0},clip,width=\columnwidth]{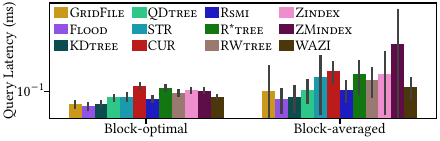}}
  \caption{
  Query latency with optimized block sizes versus expected latency over random block sizes.
  Most indexes gain $1.2$--$1.6\times$ from tuning, while \zm is the most sensitive at about $2.8\times$.
  }
  \Description{A bar chart comparing each index's tuned query latency with its expected latency over randomly selected block sizes, highlighting larger tuning sensitivity for \zm than for most other indexes.}
  \label{fig:optblocksize_vs_average}
  \vspace{-3pt}
\end{figure}

To address "\emph{RQ1: How does block size influence query latency across indexes, and which configurations yield optimal performance under varying query selectivities?}", we analyze the impact of block size on query latency.
By varying block sizes, we identify optimal configurations for each index and examine how these optima change with query selectivity.
Figure~\ref{fig:best_blocksize_plus_effect_of_selectivity} (top) illustrates the average query latency for different index structures across varying block sizes under a workload with 0.1024\% selectivity.
The latencies are averaged across multiple configurations of data and query skew.
Most latency curves exhibit a clear U-shaped pattern as block size changes, indicating a workload-dependent performance sweet spot.
This sweet spot reflects the trade-off between higher refinement overhead for many small blocks and higher scan cost when coarse blocks contain more irrelevant points.
At the default selectivity, \flood achieves the lowest tuned latency, followed closely by \grid and \kd.
The advantage of \flood over \grid is modest after tuning, but it is more robust to suboptimal block-size choices.
\qd remains slower than \kd after block-size tuning, suggesting that the block-access objective used in \cite{Yang_2020_QdtreeLearningDataa} does not fully translate to lower local query latency.
Among data-partitioning indexes, \rsmi provides the best tuned performance, while \str is close and \cur, \rw, and \rstar are slower on average.
This indicates that learned data partitioning can help, but only when its construction cost is paired with a layout that also controls scan overhead.
The query-aware data-partitioning variants do not uniformly dominate their query-agnostic counterparts.
In particular, \cur tends to choose fine blocks but pays high refinement cost, while \rw improves over \rstar in several tuned configurations without becoming a top performer.
For order-based indexes, \wazi is strongest at finer granularities, whereas \zm becomes competitive only when the block size is tuned carefully.
This makes \zm the clearest example of block-size sensitivity: its tuned latency is comparable to \zindex and \wazi, but its average latency over random block sizes is much worse.
These observations indicate that the performance of index structures can vary significantly depending on the chosen block size.
This suggests that using a fixed block size may introduce bias into performance analysis of indexing methods.
To understand the effect of using a suboptimal block size, we compare the query latency of indexes evaluated with their respective optimal block sizes against the average latency across all considered block sizes.
Figure~\ref{fig:optblocksize_vs_average} compares the query latency for indexes using optimized block sizes versus the expected latency over a random block size.
The results show that both relative and absolute performance change substantially when the block-size parameter is chosen optimally.
For example, \zm is comparable to \zindex under tuned block sizes at 0.1024\% selectivity.
However, averaging across all tested block sizes makes \zm roughly $2.8\times$ slower than its tuned latency, making it the most sensitive method in Figure~\ref{fig:optblocksize_vs_average}.

To remove the effects of block-size selection from further evaluations, we use only the results corresponding to the optimal block size for each index under each workload.
This reflects the best-case performance achievable by an administrator performing index tuning.
Figure~\ref{fig:best_blocksize_plus_effect_of_selectivity} (bottom) plots the optimal block size as a function of query selectivity, averaged across other workload parameters.
For most indexes, optimal block size increases with selectivity: highly selective queries favor finer blocks, while less selective queries tolerate coarser layouts.
The exception is \zm, whose optimal block size remains comparatively large and does not follow the same monotonic trend.
Query awareness does not imply uniformly smaller blocks: \cur and \wazi favor finer layouts than \str and \zindex, whereas \flood often uses larger average blocks than \grid while still improving latency.
These results indicate two critical insights.
First, optimizing block size is essential for a fair comparison of spatial indexes.
Second, the optimal block size is itself an informative workload signal: it reveals whether an index gains more from reducing refinement cost or from reducing false-positive scans.

\subsection{Effect of Data and Query Skew}\label{subsec:effect_data_query_skew}
\begin{figure*}[t]
        \captionsetup{skip=2pt}
  \centering
  \includegraphics[trim={0 0.05cm 0.5cm 0cm},clip,width=0.85\textwidth]{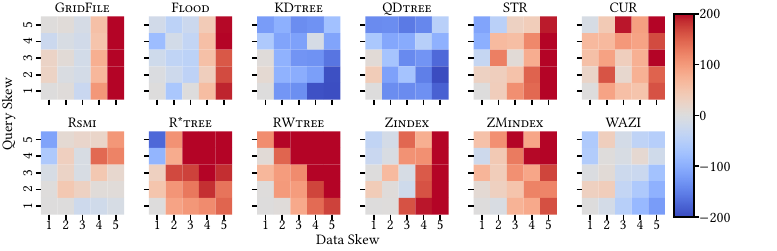}
  \caption{
        Percentage change in index performance across different levels of data skew (x-axis) and query skew (y-axis).
        The heatmap shows how each index's performance changes relative to its latency on near-uniform data and query workloads (bottom-left cell).
        Increasing data skew degrades \grid, \flood, \str, \cur, \rstar, \rw, \zindex, and \zm, while \kd, \qd, and \wazi improve or remain stable.
        Query skew has a smaller and more mixed effect: \str, \rstar, and \qd benefit in several settings, while \rw, \zm, and \cur can become slower.}
  \Description{A grid of heatmaps, one per index, with data skew on the horizontal axis and query skew on the vertical axis. Cell colors encode percentage latency change relative to the near-uniform workload.}
  \label{fig:dataskew_queryskew}
  \vspace{-3pt}
\end{figure*}

Having established the role of block size, we turn to "\emph{RQ2: How do skewed data and query distributions affect the performance of learned indexes?}".
Learned indexes improve on their classical counterparts by tailoring the index for a given data and query distribution.
Thus, the performance of learned indexes is influenced by the properties of data and query workloads.
To assess how data and query workload skew affect index performance, we analyze the relative performance changes for each index based on these two properties.
Figure~\ref{fig:dataskew_queryskew} shows how the query latency of each index changes as we vary data and query skew.
Each cell is computed as the relative percentage change in query latency for an index under a given configuration, compared with its latency under the most uniform configuration (data skew = query skew = 1, bottom left of each figure).
We observe two high-level trends in the figure.
First, data skew has a stronger and more index-specific effect than query skew.
As data skew increases, \grid, \flood, \str, \cur, \rstar, \rw, \zindex, and \zm become slower on average, with \zindex and \str showing the largest increases.
By contrast, \kd, \qd, and \wazi benefit from higher data skew, and \rsmi remains comparatively stable.
Second, query skew does not uniformly reduce latency.
Concentrating queries can reduce work for \str, \rstar, and \qd in some low-data-skew settings, but it can also increase latency for \rw, \zm, and \cur when hot query regions overlap dense or inefficient blocks.
Overall, the magnitude of the data-skew effect is larger than the query-skew effect for most indexes.
This indicates that index performance is more sensitive to the data layout induced by skew than to query concentration alone.
A skewed query workload helps only when the index layout isolates the queried dense regions into compact blocks.
The new results therefore show that learned or query-aware construction is not automatically robust to skew; its benefit depends on whether the resulting partitioning reduces false-positive scans.

\subsection{Refinement Latency versus Scan Latency} \label{subsec:refinement_vs_scan.}

\begin{figure*}
         \captionsetup{skip=3pt}
\centering
{\includegraphics[trim={0 1.7cm 0 0},clip,width=0.9\textwidth]{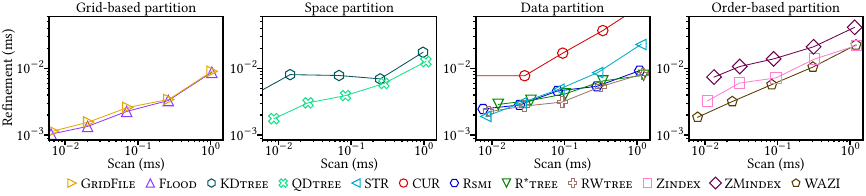}}
\caption{
        Refinement (y-axis) and scan (x-axis) latency across queries of increasing selectivity.
        Except at the smallest selectivities for a few methods, scan latency dominates refinement latency, and the gap widens rapidly as selectivity increases.
        }
\Description{A grouped latency plot with refinement time on the vertical axis and scan time on the horizontal axis. Points or curves for the evaluated indexes move primarily rightward as selectivity increases, showing scan latency growing faster than refinement latency.}
\label{fig:refinement_scan_latency_single}
  \vspace{-3pt}

\end{figure*}

While skew captures workload variability, it is also critical to separate latency components.
To answer "\emph{RQ3: How do indexes balance refinement and scan phases, and which structural designs favor one over the other?}", we measure refinement and scan latencies independently.
This reveals how indexes govern the balance between locating blocks and minimizing unnecessary scans.

Figure~\ref{fig:refinement_scan_latency_single} plots how the average scan and refinement latencies change with increasing query selectivity.
All indexes typically spend more time scanning data points than refining candidate blocks.
As selectivity increases, scan latency becomes increasingly dominant: the scan-to-refinement ratio grows from single digits at the lowest selectivity to tens or hundreds at the highest selectivity for most indexes.
\flood and \grid have very similar refinement behavior, but \flood achieves slightly lower scan cost and therefore the best tuned latency at the default selectivity.
\kd is competitive because its scan cost per result point is the lowest across most selectivity levels, even though its refinement cost is higher than the grid-based indexes.
\qd reduces refinement cost relative to \kd, but it scans more points and is therefore slower after tuning.
This indicates that minimizing block accesses alone is insufficient when the selected blocks contain more false-positive points.
\rsmi is the best data-partitioning method after tuning, with lower total latency than \cur, \rw, and \rstar and latency close to \str.
The iterative-insertion indexes \rstar and \rw do not consistently reduce scan cost enough to offset their construction and layout complexity.
\cur is the clearest outlier in this family: it uses fine blocks, but its refinement overhead is high.
The order-based indexes show a different trade-off.
\wazi minimizes false positives, especially at higher selectivities, but \kd, \flood, and \grid still achieve lower scan time per result point.
\zindex and \zm scan more false positives than \wazi; \zm also has the highest refinement overhead among the order-based methods.
Consequently, \zm is not the high-selectivity winner in the new synthetic results despite being competitive under tuned block sizes in Section~\ref{subsec:optimal_block_sizes}.
The high-selectivity advantage instead belongs to indexes that combine low scan cost per result point with controlled false positives.

\begin{figure*}
        \captionsetup{skip=2pt}
\centering
{\begin{minipage}[b]{0.45\textwidth}
        \includegraphics[trim={0 0cm 0 0cm},clip,width=\textwidth]{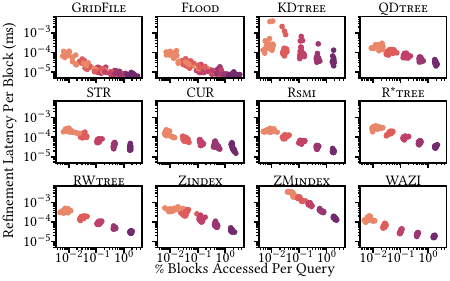}
\end{minipage}
\qquad
\begin{minipage}[b]{0.45\textwidth}
        {\includegraphics[trim={0 0cm 0 0cm},clip,width=\textwidth]{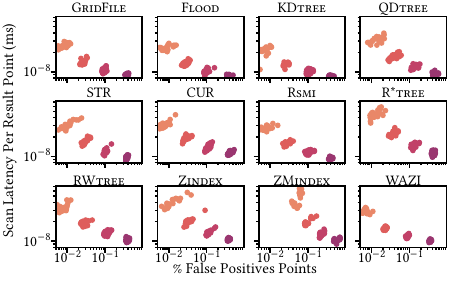}}
\end{minipage}}
{\begin{minipage}[b]{\textwidth}
        {\includegraphics[trim={0 0.1cm 0 0cm},clip,width=\textwidth]{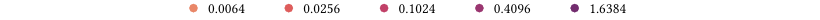}}
\end{minipage}}
\caption{
        Left: refinement latency per block versus blocks accessed. Right: scan latency per result point versus false-positive points scanned. Color encodes query selectivity.
        False positives and scan latency do not perfectly align: \wazi minimizes false positives, while \kd, \flood, and \grid scan fastest.
        }
\Description{Two scatter plots with a shared color legend for query selectivity. The left plot relates the fraction of blocks accessed to refinement latency per block, and the right plot relates false-positive scans to scan latency per result point.}
\label{fig:PercentBlocksAccessed_RefinementLatency_PercentBlocksAccessed_ScanLatency}
\vspace{-3pt}
\end{figure*}

\begin{figure*}[ht!]
        \captionsetup{skip=2pt}
      \centering
      {\includegraphics[trim={0 0.2cm 0 0},clip,width=0.9\textwidth]{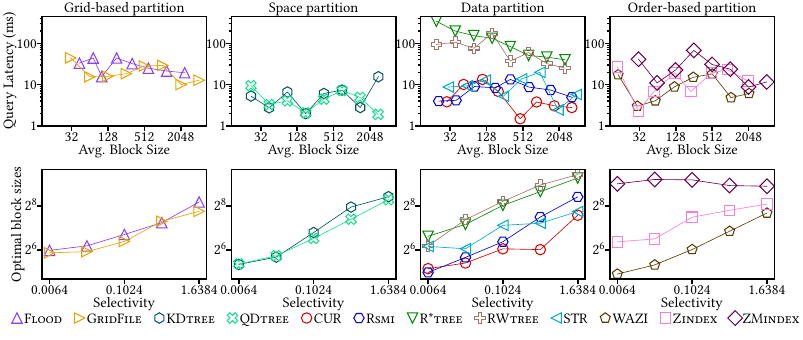}}
      \caption{
        Performance comparison of indexes on disk-backed data pages.
        Very fine blocks often incur high disk-backed latency, but the best block size remains index- and workload-dependent rather than simply favoring the coarsest layout.
        Optimal block size generally still increases with query selectivity, although the disk-backed optima are not uniformly larger than the in-memory optima.
        }
      \Description{A two-part disk-backed performance plot. The upper panels show query latency across block sizes for each index, and the lower panels show how the optimal disk-backed block size changes across query selectivities.}
      \label{fig:diskbacked_best_blocksize_plus_effect_of_selectivity}
        \vspace{-3pt}
\end{figure*}

Figure~\ref{fig:PercentBlocksAccessed_RefinementLatency_PercentBlocksAccessed_ScanLatency} (left) shows the changes in refinement latency per block and the percentage of blocks accessed for each index over queries of different selectivity.
The left plot shows that block-access behavior is strongly family-dependent.
\grid and \flood access similar fractions of blocks and have low per-block refinement costs.
\kd and \qd also have similar access patterns, but \qd's selected blocks contain more irrelevant points, leading to worse scan behavior.
Among data-partitioning methods, \cur pays the highest refinement overhead, whereas \rsmi and \rw access moderate fractions of blocks with lower per-block refinement cost.
Thus, the learned variants do not all behave the same: \flood and \qd reduce refinement overhead relative to their classical counterparts, while \cur and \zm are refinement-heavy.
The spread within each selectivity level also confirms the finding from Section~\ref{subsec:effect_data_query_skew}: data and query skew affect how many blocks a query must refine.

The scan latency of a query is directly correlated with the number of points scanned for each index.
The differences in scan latency between indexes depend on the number of false-positive points scanned during query processing.
Figure~\ref{fig:PercentBlocksAccessed_RefinementLatency_PercentBlocksAccessed_ScanLatency} (right) shows the changes in scan latency per result point and false positives for each index over queries of different selectivity.
We make four observations from the figure.
First, \wazi consistently scans the fewest false-positive points among the order-based indexes and is often the best overall on this metric.
Second, low false-positive counts do not fully determine scan latency: \kd, \flood, and \grid usually have the lowest scan latency per result point.
Third, \rstar remains weak among data-partitioning indexes, while \rw improves over \rstar but rarely reaches the top tier.
\zm incurs more false positives than \wazi and combines that with higher refinement cost, explaining its sensitivity to block-size tuning.
Finally, the best scan efficiency is achieved by simple layouts with low per-point scan overhead, especially \kd, \flood, and \grid.

Across all indexes, scan latency dominates once selectivity grows, but the source of scan efficiency differs by family. \wazi is strongest at minimizing false positives, whereas \kd, \flood, and \grid provide the lowest scan latency per result point. In contrast, \rstar, \cur, and \zm show that sophisticated partitioning can still lose when it creates either high refinement overhead or high false-positive scans.

\subsection{Considerations of Indexing Disk-Based Data} \label{subsec:disk_based_data}

The previous analysis assumed in-memory configurations.
To address "\emph{RQ4: How do disk-based storage conditions alter the optimal block size and the refinement-scan latency trade-off compared to in-memory settings?}", we extend evaluation to disk-backed data.
This highlights how slower access costs shift block size preferences and alter refinement-to-scan trade-offs.
The optimal block sizes presented in Section~\ref{subsec:optimal_block_sizes} identify the sweet spot between added refinement cost (more index layers and smaller block sizes) and reduced scan time from smaller block sizes.
This trade-off is determined by the latency of data access during scanning, which remains nearly constant in in-memory settings but changes substantially on disk due to slower access.
To analyze this effect, we store data blocks in a memory-mapped file and evaluate index performance across all data and query configurations under disk-backed conditions.

\begin{figure}[t]
        \captionsetup{skip=2pt}
        \centering
        {\includegraphics[trim={0 0.0cm 0 0cm},clip,width=0.75\columnwidth]{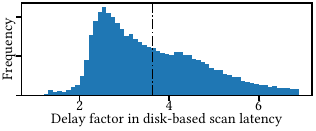}}
        \caption{
                Histogram of latency-change factor between disk-backed and in-memory configurations.
                On average, scanning disk-backed data was about 3.6$\times$ (dotted line) slower.
        }
        \Description{A histogram of latency-change factors comparing disk-backed and in-memory configurations, with a dotted vertical line marking the average slowdown of about 3.6 times.}
        \label{fig:DiskBackedDelay}
        \vspace{-3pt}
\end{figure}

Our experiments show that the effective data-access cost of disk-backed scans is about $3.6\times$ higher than in-memory scans, as shown in Figure~\ref{fig:DiskBackedDelay}.
We call this the effective data-access cost because the actual cost of accessing a block from disk is at least $100\times$ higher than accessing a similar block from memory.
The access cost is lowered because the operating system caches recently accessed blocks, reducing measured latency.
Figure~\ref{fig:diskbacked_best_blocksize_plus_effect_of_selectivity} (top) illustrates the average query latency for different index structures across varying block sizes under a workload with 0.1024\% selectivity.
Unlike the in-memory results, the disk-backed curves are less uniformly U-shaped.
Very small blocks can be expensive because they require many page fetches, but increasing block size eventually raises scan cost; the balance point differs by index family.
Figure~\ref{fig:diskbacked_best_blocksize_plus_effect_of_selectivity} (bottom) shows that optimal block size generally increases with selectivity, but the absolute values are often smaller than their in-memory counterparts because disk-backed evaluation penalizes excessive scans.
At the default selectivity, the fastest disk-backed methods are \kd, \rsmi, \wazi, and \qd, indicating that structural differences still matter under higher access cost.
Overall, these results demonstrate that the disk-backed cost model differs qualitatively from the in-memory setting: it penalizes both excessive block fetches and excessive false-positive scans.

We identify two key insights from these results.
First, disk-backed performance cannot be predicted from block size alone: \flood is faster than \grid at the default selectivity, but the relative block sizes of the two indexes vary across workloads.
Second, query selectivity remains the strongest driver of the optimal granularity, while the index structure determines whether that granularity translates into low scan cost.

\subsection{Index Construction Cost}

\begin{table*}[ht!]
        \captionsetup{skip=2pt}
\caption{
        Build times of each index (seconds) and their redemption relative to the \cur index (gray).
        Redemption indicates the number of queries (in millions) needed for an index to amortize its build cost compared with \cur.
        }
\Description{A table showing build time and redemption thresholds for each evaluated index in both in-memory and disk-backed settings, with the \cur baseline column shaded.}
    \centering
    \setlength{\tabcolsep}{2pt}
    \begin{tabular}{c|cc|cc|c>{\columncolor[gray]{0.8}}cccc|ccc}
        \toprule
        Index                   &                \grid &               \flood &              \kd &             \qd &           \str &   \cur      &    \rsmi      &   \rstar &         \rw &     \zindex &         \zm &   \wazi \\ \midrule
        Build Time (sec)        &                  9.4 &                148.6 &             16.4 &            24.1 &            6.5 &   13.6      &     2031.2   &      38.1 &        44.6 &        19.8 &        23.5 &   38.2 \\
        Redemption (in-memory)  &                  (+) &           (+)$8.9M$ &        (+)$0.15M$ &      (+)$0.44M$ &            (+) &             &  (+)$790M$   &       (-) &    (+)$21M$ &    (+)$15M$ &  (+)$0.48M$ &  (+)$2.6M$ \\
        Redemption (disk-based) &           (-)$0.26M$ &                (-) &          (+)$1.2M$ &       (+)$0.2M$ &     (-)$0.13M$ &             &  (+)$4.1M$   &       (-) &         (-) &         (-) &         (-) &  (+)$0.29M$ \\
        \bottomrule
    \end{tabular}
\label{tab:build_time_redemption}
\vspace{-3pt}
\end{table*}

Beyond latency, index adoption also depends on relative construction overhead.
In response to "\emph{RQ5: What are the construction costs of different indexes, and under what query volumes are these costs effectively amortized?}", we measure build times and compute redemption thresholds.
These results indicate when higher upfront construction cost is offset by long-term query savings.
Table~\ref{tab:build_time_redemption} reports build times for all indexes.
We observe that learned indexes generally increase construction cost, but the magnitude varies substantially by family.
\str and \grid have the lowest construction times, as their iterative sort-based algorithms have log-linear complexity.
\grid sorts the data twice, while \str repeatedly sorts until blocks of the required granularity are achieved.
The layout optimization in \flood is expensive, taking about $16\times$ longer than \grid, whereas \qd is only about $1.5\times$ slower to build than \kd.
Among data-partitioning indexes, \rsmi is the clear construction-cost outlier, while \cur and \rw add only moderate overhead relative to \str and \rstar.
Order-based learned methods remain comparatively cheap: \zm and \wazi are constant-factor slower than \zindex.
We are also interested in the cost redemption of each index.
We define cost redemption, measured relative to the query-aware \cur baseline, as the number of queries for an index to amortize its construction cost: $redem = \mathcal{I}^{+/-} \cdot \frac{index\ build\ time - base\ build\ time}{base\ query\ latency-index\ query\ latency}$, where $\mathcal{I}^{+/-}$ sets the table sign.
In Table~\ref{tab:build_time_redemption}, (+) denotes eventual redemption and (-) no long-run redemption; the attached number is the crossover query count, and a missing number means the index is always or never redeemed.
We observe two key findings from the redemption behavior.
First, in the in-memory setting, all indexes except \rstar eventually redeem their construction cost relative to \cur.
The redemption threshold, however, varies widely: \kd, \qd, and \zm redeem within fewer than half a million queries, whereas \rsmi requires roughly $790$M queries because of its very high build time.
Second, disk-backed redemption changes the picture: \kd, \qd, \rsmi, and \wazi redeem their construction costs, while \flood, \rstar, \rw, \zindex, and \zm do not.
Low-build-time indexes such as \grid and \str can still be preferable for small disk-backed query volumes, but their slower disk query latency prevents long-run redemption.

\subsection{Index Selection} \label{subsec:index_selection}

Query performance varies widely across workloads.
In our experiments, four factors primarily drive these differences: data skew, query skew, storage medium, and selectivity.
We therefore answer "\emph{RQ6: Given data and query workload statistics, which index is expected to provide the best range query performance?}" by learning a simple decision tree that maps these factors to an index choice.

\begin{figure}[t]
        \captionsetup{skip=2pt}
        \centering
        \begin{tikzpicture}[level distance=1.6cm
                , edge from parent path={(\tikzparentnode) |- (\tikzchildnode)}
                ]

                \node[align=center] (root){Disk-backed\\Data?}[grow=0]
                { [sibling distance=3.45cm,level distance=1.2cm]
                        child { node[align=center] {query skew $\leq 3$}
                                {[sibling distance=2cm]
                                        child {node[align=center] {data skew $\leq 3$}
                                        {[sibling distance=1cm,level distance=2cm]
                                                        child {node {$\rw/\flood/\grid$}}
                                                        child {node {$\wazi/\rsmi/\cur$}}
                                                }
                                        }
                                        child {node[align=center] {data skew $\leq 3$}
                                        {[sibling distance=1cm,level distance=2cm]
                                                        child {node {$\zindex/\qd/\str$}}
                                                        child {node {$\wazi/\cur/\kd$}}
                                                }
                                        }
                                }
                        }
                        child {node[align=center] {data skew $\leq 3$}
                                {[sibling distance=1.125cm,level distance=2cm]
                                        child {node {selectivity  $\leq 0.0256\%$}
                                                {[sibling distance=1cm,level distance=2cm]
                                                        child {node {$\grid/\flood/\kd $}}
                                                        child {node {$\str/\qd/\rsmi$}}
                                                }
                                        }
                                        child {node {$\wazi/\kd/\flood $}}
                                }
                        }
                };

                        \node[above=0.7cm of root,anchor=center,rotate=90,align=center] (nonode) {No \\ };
                        \node[below=0.7cm of root,anchor=center,rotate=90,align=center] (yesnode) {Yes \\ };

          \end{tikzpicture}
          \caption{
                Decision tree classifier trained to select one of the top-performing indexes based on data skew, query skew, selectivity, and storage medium.
                The tree predicts at least one top-three index with 57.8\% overall top-three accuracy.
                }
          \Description{A decision tree whose root tests whether data is disk-backed. The branches split on query skew, data skew, and selectivity, and the leaves list three recommended indexes for each workload region.}
          \label{fig:decision_tree}
          \vspace{-3pt}
      \end{figure}
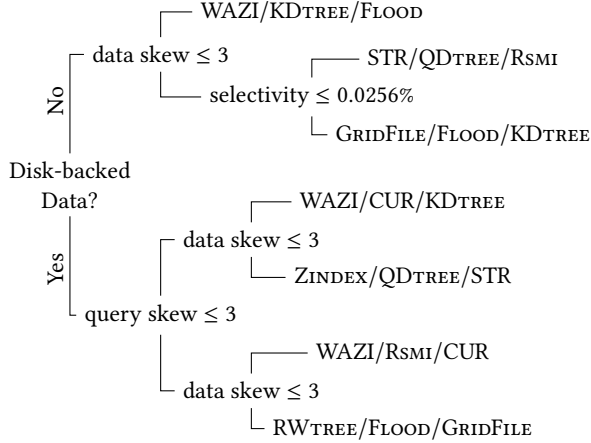

To construct the decision tree, we rank indexes by query latency for each data sample and query workload.
For each measurable configuration, defined by data skew, query skew, selectivity, and storage medium, we retain the top-three indexes as candidate labels, so the classifier learns regions where an index is competitive rather than only where it is the single fastest method.
The resulting simplified tree in Figure~\ref{fig:decision_tree} predicts at least one top-three candidate with 57.8\% overall top-three accuracy.

In the in-memory setting, \flood, \kd, and \grid account for nearly all top-three outcomes.
\flood dominates low and medium selectivities, while \kd becomes increasingly competitive at high selectivity; \grid remains a strong baseline across the range.
For disk-backed data, the top-three set is broader: \kd appears most often, followed by \rsmi, \flood, \grid, and \wazi.
At high disk-backed selectivity, \rsmi, \flood, \kd, and \grid are most frequent among the top-three choices, while \wazi and \qd are useful in more specialized regions.

\subsection{Validation of Synthetic-Data Insights}\label{subsec:validation_synthetic_insights}

To answer "\emph{RQ7: Do the index-selection insights learned from synthetic data generalize to real-world data distributions?}", we evaluate the decision tree derived from the synthetic experiments on held-out real-world validation settings.
Each validation instance contains 8 million two-dimensional point locations sampled from OpenStreetMap, providing spatial layouts generated by real geographic processes rather than by the synthetic GMM data generator.
We pair each OpenStreetMap point set with 1,000 synthetic Gaussian-mixture queries with randomly selected query-skew levels and selectivities, testing whether the selection rules learned from synthetic data transfer to real spatial point distributions.
To validate the synthetic-study results, we analyze how effectively the decision tree selects the best-performing indexes on real-world spatial data.
We generate 25 validation datasets from OpenStreetMap, each with 8 million points, and pair each dataset with 1,000 synthetic queries.
For each dataset--workload pair, we build every index separately using the block-size settings identified by the synthetic experiments for the in-memory and disk-backed cases.
For practical index selection, the relevant question is not only whether the tree names the exact fastest index, but how much latency is lost by following its recommendations.
We therefore measure top-three decision regret: the latency of the best index among the tree's three recommendations divided by the latency of the empirical optimum.
The in-memory results show strong transfer: the decision tree places the best-performing index in its top-three recommendations for 72\% of the OpenStreetMap scenarios, closely matching its 70.9\% in-memory top-three accuracy on the synthetic configurations.
More importantly, the median top-three decision regret is $1.0\times$ in memory, and the best recommended index is within $1.25\times$ of the observed optimum in every in-memory validation scenario.
Disk-backed validation is harder because block-layout errors are amplified by I/O cost.
Although the top-three hit rate drops to 48\%, the recommendations still have low practical cost: the median top-three decision regret is only $1.008\times$, and in 84\% of scenarios the top-three set includes an index within $1.25\times$ of the best disk-backed latency.

These results show that the controlled synthetic study captures transferable structure in the index-selection problem: the learned rules do not merely fit synthetic distributions, but provide actionable recommendations that typically recover the best index or one with nearly identical latency on real OpenStreetMap layouts.

\section{Conclusions} \label{sec:conclusion}

This study shows that learned spatial indexes can outperform traditional indexes, but their advantage depends on specific conditions such as data characteristics and workload. 
Using a unified benchmarking framework, we compared multiple learned spatial indexes against established spatial indexing techniques. 
By optimizing index granularity for each workload, we evaluated the indexes near their strongest observed configurations and avoided conclusions that depend on a single fixed block size. 
The results demonstrate that learned spatial indexes are effective, although their advantage is conditional rather than universal.
Block-size tuning emerges as a primary factor; most indexes improve substantially after tuning, and the optimal granularity varies with selectivity, mediating the trade-off between refinement overhead and false-positive scans.
Across synthetic workloads, learned and query-aware layouts are most effective when they reduce scan costs without incurring excessive refinement costs.
This finding accounts for the consistent in-memory performance of \flood, \kd, and \grid, the prominence of \rsmi, \wazi, and \qd under disk-backed evaluation, and the underperformance of methods with sophisticated construction when their block layout results in numerous false positives.
The experiments also show that data skew is often more consequential than query skew, and that disk-backed storage changes the cost model by penalizing both excessive page accesses and unnecessary scans.
Construction costs further qualify these findings; several learned indexes amortize their build overhead only after sufficient query volume, indicating that deployment decisions should account for both query latency and amortization.
These findings are synthesized into a decision tree that uses workload statistics to recommend competitive indexes.
On real OpenStreetMap validation workloads, exact winner recovery is imperfect, especially on disk-backed data, but the resulting decision regret is minimal: the median top-three regret is $1.0\times$ in memory and $1.008\times$ on disk-backed data.
This means that the decision tree usually recommends either the best index or one with nearly identical latency, making the learned selection rules useful as practical guidance.
Collectively, these results underscore the value of this study as both an empirical benchmark and a practical guide for selecting spatial indexes under explicit workload and storage assumptions.

\subsubsection*{Limitations and Future Work}
Synthetic datasets allowed precise control of data and query skew, but broader validation should include diverse real-world datasets.
Our analysis focuses on static data; dynamic environments~\cite{Gu_2023_RLRTreeReinforcementLearningBased}, including adaptive retraining~\cite{Hidaka2024FlexFloodEU} and incremental learning, remain important extensions.
Beyond the representative methods studied here, further variations and hybrid designs need systematic exploration to better understand trade-offs between complexity, training cost, and efficiency.
Expanding the framework to support query types beyond range queries, such as nearest-neighbor search~\cite{Yongxin_2020_StudyLearnedKDa} and spatial joins, would broaden its practical scope.
Because we simplified some implementations for comparability, validating results against production-grade systems and heterogeneous computing environments is an important next step.

\clearpage

\section*{Artifacts}
The code developed for this work is publicly available at \gitlabrepo.
The repository contains the Python script needed for \rsmi training and data generation, C++ implementations of all evaluated indexes, experiment configuration files, and scripts used to produce the reported measurements.
Its README describes the required dependencies and setup process, provides a small smoke test for checking that the pipeline builds and runs locally, and gives the commands for reproducing the full synthetic experiments and for running the same workflow through Slurm.
Real-workload preprocessing instructions are also included, so the artifact documents both quick validation and full experimental reproduction.

\bibliographystyle{ACM-Reference-Format}
\bibliography{references}

@inproceedings{Kraska_2018_CaseLearnedIndexa,
  title = {The {{Case}} for {{Learned Index Structures}}},
  booktitle = {Proceedings of the 2018 {{International Conference}} on {{Management}} of {{Data}}},
  author = {Kraska, Tim and Beutel, Alex and Chi, Ed H. and Dean, Jeffrey and Polyzotis, Neoklis},
  year = {2018},
  month = may,
  series = {{{SIGMOD}} '18},
  pages = {489--504},
  publisher = {Association for Computing Machinery},
  address = {New York, NY, USA},
  doi = {10.1145/3183713.3196909},
  isbn = {978-1-4503-4703-7},
}

@article{Ferragina_2020_PGMindexFullydynamicCompressed,
  title = {The {{PGM-index}}: A Fully-Dynamic Compressed Learned Index with Provable Worst-Case Bounds},
  author = {Ferragina, Paolo and Vinciguerra, Giorgio},
  year = {2020},
  month = apr,
  journal = {Proceedings of the VLDB Endowment},
  volume = {13},
  number = {8},
  pages = {1162--1175},
  issn = {2150-8097},
  doi = {10.14778/3389133.3389135},
}

@article{Abu-Libdeh_2020_LearnedIndexesGooglescaleDiskbased,
  author       = {Hussam Abu{-}Libdeh and
                  Deniz Altinb{\"{u}}ken and
                  Alex Beutel and
                  Ed H. Chi and
                  Lyric Doshi and
                  Tim Kraska and
                  Xiaozhou Li and
                  Andy Ly and
                  Christopher Olston},
  title        = {Learned Indexes for a Google-scale Disk-based Database},
  journal      = {CoRR},
  volume       = {abs/2012.12501},
  year         = {2020},
  url          = {https://arxiv.org/abs/2012.12501},
  eprinttype   = {arXiv},
  eprint       = {2012.12501},
}

@inproceedings{DBLP:conf/eci/NievergeltHS81,
        author       = {J{\"{u}}rg Nievergelt and
                        Hans Hinterberger and
                        Kenneth C. Sevcik},
        editor       = {A. J. W. Duijvestijn and
                        Peter C. Lockemann},
        title        = {The Grid File: An Adaptable, Symmetric Multi-Key File Structure},
        booktitle    = {Trends in Information Processing Systems, 3rd Conference of the European
                        Cooperation in Informatics, Munich, Germany, October 20-22, 1981,
                        Proceedings},
        series       = {Lecture Notes in Computer Science},
        volume       = {123},
        pages        = {236--251},
        publisher    = {Springer},
        year         = {1981},
        url          = {https://doi.org/10.1007/3-540-10885-8\_45},
        doi          = {10.1007/3-540-10885-8\_45},
        }

@inproceedings{DBLP:conf/sigmod/NathanDAK20,
        author       = {Vikram Nathan and
                        Jialin Ding and
                        Mohammad Alizadeh and
                        Tim Kraska},
        editor       = {David Maier and
                        Rachel Pottinger and
                        AnHai Doan and
                        Wang{-}Chiew Tan and
                        Abdussalam Alawini and
                        Hung Q. Ngo},
        title        = {Learning Multi-Dimensional Indexes},
        booktitle    = {Proceedings of the 2020 International Conference on Management of
                        Data, {SIGMOD} Conference 2020, online conference [Portland, OR, USA],
                        June 14-19, 2020},
        pages        = {985--1000},
        publisher    = {{ACM}},
        year         = {2020},
        url          = {https://doi.org/10.1145/3318464.3380579},
        doi          = {10.1145/3318464.3380579},
        }

@inproceedings{DBLP:conf/sigmod/Li0ZY020,
        author       = {Pengfei Li and
                        Hua Lu and
                        Qian Zheng and
                        Long Yang and
                        Gang Pan},
        editor       = {David Maier and
                        Rachel Pottinger and
                        AnHai Doan and
                        Wang{-}Chiew Tan and
                        Abdussalam Alawini and
                        Hung Q. Ngo},
        title        = {{LISA:} {A} Learned Index Structure for Spatial Data},
        booktitle    = {Proceedings of the 2020 International Conference on Management of
                        Data, {SIGMOD} Conference 2020, online conference [Portland, OR, USA],
                        June 14-19, 2020},
        pages        = {2119--2133},
        publisher    = {{ACM}},
        year         = {2020},
        url          = {https://doi.org/10.1145/3318464.3389703},
        doi          = {10.1145/3318464.3389703},
    }

@article{Hidaka2024FlexFloodEU,
    author       = {Fuma Hidaka and
                  Yusuke Matsui},
  title        = {FlexFlood: Efficiently Updatable Learned Multi-dimensional Index},
  journal      = {CoRR},
  volume       = {abs/2411.09205},
  year         = {2024},
  url          = {https://doi.org/10.48550/arXiv.2411.09205},
  doi          = {10.48550/ARXIV.2411.09205},
  eprinttype   = {arXiv},
  eprint       = {2411.09205},
}

@article{Bentley_1975_MultidimensionalBinarySearch,
        title = {Multidimensional Binary Search Trees Used for Associative Searching},
        author = {Bentley, Jon Louis},
        year = {1975},
        month = sep,
        journal = {Commun. ACM},
        volume = {18},
        number = {9},
        pages = {509--517},
        issn = {0001-0782},
        doi = {10.1145/361002.361007},
        }

@inproceedings{Yang_2020_QdtreeLearningDataa,
        title = {Qd-Tree: {{Learning Data Layouts}} for {{Big Data Analytics}}},
        booktitle = {Proceedings of the 2020 {{ACM SIGMOD International Conference}} on {{Management}} of {{Data}}},
        author = {Yang, Zongheng and Chandramouli, Badrish and Wang, Chi and Gehrke, Johannes and Li, Yinan and Minhas, Umar Farooq and Larson, Per-{\AA}ke and Kossmann, Donald and Acharya, Rajeev},
        year = {2020},
        month = jun,
        pages = {193--208},
        publisher = {ACM},
        address = {Portland OR USA},
        doi = {10.1145/3318464.3389770},
        isbn = {978-1-4503-6735-6},
    }

@inproceedings{Yongxin_2020_StudyLearnedKDa,
        title = {A {{Study}} of {{Learned KD Tree Based}} on {{Learned Index}}},
        booktitle = {2020 {{International Conference}} on {{Networking}} and {{Network Applications}} ({{NaNA}})},
        author = {Yongxin, Peng and Wei, Zhou and Lin, Zhang and Hongle, Du},
        year = {2020},
        month = dec,
        pages = {355--360},
        doi = {10.1109/NaNA51271.2020.00067},
    }

@inproceedings{DBLP:conf/sigmod/Guttman84,
        author       = {Antonin Guttman},
        editor       = {Beatrice Yormark},
        title        = {R-Trees: {A} Dynamic Index Structure for Spatial Searching},
        booktitle    = {SIGMOD'84, Proceedings of Annual Meeting, Boston, Massachusetts, USA,
                        June 18-21, 1984},
        pages        = {47--57},
        publisher    = {{ACM} Press},
        year         = {1984},
        url          = {https://doi.org/10.1145/602259.602266},
        doi          = {10.1145/602259.602266},
    }

@inproceedings{DBLP:conf/sigmod/BeckmannKSS90,
        author       = {Norbert Beckmann and
                        Hans{-}Peter Kriegel and
                        Ralf Schneider and
                        Bernhard Seeger},
        editor       = {Hector Garcia{-}Molina and
                        H. V. Jagadish},
        title        = {The R*-Tree: An Efficient and Robust Access Method for Points and
                        Rectangles},
        booktitle    = {Proceedings of the 1990 {ACM} {SIGMOD} International Conference on
                        Management of Data, Atlantic City, NJ, USA, May 23-25, 1990},
        pages        = {322--331},
        publisher    = {{ACM} Press},
        year         = {1990},
        url          = {https://doi.org/10.1145/93597.98741},
        doi          = {10.1145/93597.98741},
    }

@inproceedings{Beckmann_2009_RevisedRtreeComparison,
        title = {A Revised R*-Tree in Comparison with Related Index Structures},
        booktitle = {Proceedings of the 2009 {{ACM SIGMOD International Conference}} on {{Management}} of Data},
        author = {Beckmann, Norbert and Seeger, Bernhard},
        year = {2009},
        month = jun,
        pages = {799--812},
        publisher = {ACM},
        address = {Providence Rhode Island USA},
        doi = {10.1145/1559845.1559929},
        isbn = {978-1-60558-551-2},
    }

@inproceedings{Dong_2022_RWTreeLearnedWorkloadawarea,
        title = {{{RW-Tree}}: {{A Learned Workload-aware Framework}} for {{R-tree Construction}}},
        booktitle = {2022 {{IEEE}} 38th {{International Conference}} on {{Data Engineering}} ({{ICDE}})},
        author = {Dong, Haowen and Chai, Chengliang and Luo, Yuyu and Liu, Jiabin and Feng, Jianhua and Zhan, Chaoqun},
        year = {2022},
        month = may,
        pages = {2073--2085},
        issn = {2375-026X},
        doi = {10.1109/ICDE53745.2022.00201},
    }

@inproceedings{Leutenegger_1997_STRSimpleEfficienta,
        title = {{{STR}}: A Simple and Efficient Algorithm for {{R-tree}} Packing},
        booktitle = {Proceedings 13th {{International Conference}} on {{Data Engineering}}},
        author = {Leutenegger, S.T. and Lopez, M.A. and Edgington, J.},
        year = {1997},
        month = apr,
        pages = {497--506},
        issn = {1063-6382},
        doi = {10.1109/ICDE.1997.582015},
    }

@inproceedings{Ross_2001_CostbasedUnbalancedRtrees,
        title = {Cost-Based Unbalanced {{R-trees}}},
        booktitle = {Proceedings {{Thirteenth International Conference}} on {{Scientific}} and {{Statistical Database Management}}. {{SSDBM}} 2001},
        author = {Ross, K.A. and Sitzmann, I. and Stuckey, P.J.},
        year = {2001},
        month = jul,
        pages = {203--212},
        issn = {1099-3371},
        doi = {10.1109/SSDM.2001.938552},
    }

@article{Qi_2020_EffectivelyLearningSpatiala,
        title = {Effectively Learning Spatial Indices},
        author = {Qi, Jianzhong and Liu, Guanli and Jensen, Christian S. and Kulik, Lars},
        year = {2020},
        month = aug,
        journal = {Proceedings of the VLDB Endowment},
        volume = {13},
        number = {12},
        pages = {2341--2354},
        issn = {2150-8097},
        doi = {10.14778/3407790.3407829},
    }

@article{Gu_2023_RLRTreeReinforcementLearningBased,
author = {Gu, Tu and Feng, Kaiyu and Cong, Gao and Long, Cheng and Wang, Zheng and Wang, Sheng},
title = {The RLR-Tree: A Reinforcement Learning Based R-Tree for Spatial Data},
year = {2023},
issue_date = {May 2023},
publisher = {Association for Computing Machinery},
address = {New York, NY, USA},
volume = {1},
number = {1},
url = {https://doi.org/10.1145/3588917},
doi = {10.1145/3588917},
journal = {Proc. ACM Manag. Data},
month = may,
articleno = {63},
numpages = {26},
}

@article{tropf1981multidimensional,
        title={Multidimensional Range Search in Dynamically Balanced Trees.},
        author={Tropf, Herbert and Herzog, Helmut},
        journal={Angewandte Info.},
        number={2},
        pages={71--77},
        year={1981}
    }

@article{DBLP:journals/gidr/Markl00,
        author       = {Volker Markl},
        title        = {Mistral - Processing Relational Queries using a Multidimensional Access
                        Technique},
        journal      = {Datenbank Rundbr.},
        volume       = {26},
        pages        = {24--25},
        year         = {2000},
    }

@inproceedings{DBLP:conf/wwca/Bayer97,
        author       = {Rudolf Bayer},
        editor       = {Takashi Masuda and
                        Yoshifumi Masunaga and
                        Michiharu Tsukamoto},
        title        = {The Universal B-Tree for Multidimensional Indexing: General Concepts},
        booktitle    = {Worldwide Computing and Its Applications, International Conference,
                        {WWCA} '97, Tsukuba, Japan, March 10-11, 1997, Proceedings},
        series       = {Lecture Notes in Computer Science},
        volume       = {1274},
        pages        = {198--209},
        publisher    = {Springer},
        year         = {1997},
        url          = {https://doi.org/10.1007/3-540-63343-X\_48},
        doi          = {10.1007/3-540-63343-X\_48},
    }

@inproceedings{DBLP:conf/edbt/PaiM024,
  author       = {Sachith Pai and
                  Michael Mathioudakis and
                  Yanhao Wang},
  editor       = {Letizia Tanca and
                  Qiong Luo and
                  Giuseppe Polese and
                  Loredana Caruccio and
                  Xavier Oriol and
                  Donatella Firmani},
  title        = {WaZI: {A} Learned and Workload-aware Z-Index},
  booktitle    = {Proceedings 27th International Conference on Extending Database Technology,
                  {EDBT} 2024, Paestum, Italy, March 25 - March 28},
  pages        = {559--571},
  publisher    = {OpenProceedings.org},
  year         = {2024},
  url          = {https://doi.org/10.48786/edbt.2024.48},
  doi          = {10.48786/EDBT.2024.48},
}

@inproceedings{Wang_2019_LearnedIndexSpatiala,
  title = {Learned {{Index}} for {{Spatial Queries}}},
  booktitle = {2019 20th {{IEEE International Conference}} on {{Mobile Data Management}} ({{MDM}})},
  author = {Wang, Haixin and Fu, Xiaoyi and Xu, Jianliang and Lu, Hua},
  year = {2019},
  month = jun,
  pages = {569--574},
  issn = {2375-0324},
  doi = {10.1109/MDM.2019.00121},
}

@article{Kipf_2019_SOSDBenchmarkLearned,
  author       = {Andreas Kipf and
                  Ryan Marcus and
                  Alexander van Renen and
                  Mihail Stoian and
                  Alfons Kemper and
                  Tim Kraska and
                  Thomas Neumann},
  title        = {{SOSD:} {A} Benchmark for Learned Indexes},
  journal      = {CoRR},
  volume       = {abs/1911.13014},
  year         = {2019},
  url          = {http://arxiv.org/abs/1911.13014},
  eprinttype   = {arXiv},
  eprint       = {1911.13014},
}

@article{Sun__LearnedIndexComprehensive,
  author       = {Zhaoyan Sun and
                  Xuanhe Zhou and
                  Guoliang Li},
  title        = {Learned Index: {A} Comprehensive Experimental Evaluation},
  journal      = {Proc. {VLDB} Endow.},
  volume       = {16},
  number       = {8},
  pages        = {1992--2004},
  year         = {2023},
  url          = {https://www.vldb.org/pvldb/vol16/p1992-li.pdf},
  doi          = {10.14778/3594512.3594528},
  bibsource    = {dblp computer science bibliography, https://dblp.org}
}

@article{Marcus_2020_BenchmarkingLearnedIndexesa,
  title = {Benchmarking {{Learned Indexes}}},
  author = {Marcus, Ryan and Kipf, Andreas and van Renen, Alexander and Stoian, Mihail and Misra, Sanchit and Kemper, Alfons and Neumann, Thomas and Kraska, Tim},
  year = {2020},
  month = sep,
  journal = {Proceedings of the VLDB Endowment},
  volume = {14},
  number = {1},
  eprint = {2006.12804},
  pages = {1--13},
  issn = {2150-8097},
  doi = {10.14778/3421424.3421425},
}

@article{Maltry_2022_CriticalAnalysisRecursive,
  title = {A Critical Analysis of Recursive Model Indexes},
  author = {Maltry, Marcel and Dittrich, Jens},
  year = {2022},
  month = jan,
  journal = {Proceedings of the VLDB Endowment},
  volume = {15},
  number = {5},
  pages = {1079--1091},
  issn = {2150-8097},
  doi = {10.14778/3510397.3510405},
}

@article{Wongkham_2022_AreUpdatableLearneda,
  title = {Are {{Updatable Learned Indexes Ready}}?},
  author = {Wongkham, Chaichon and Lu, Baotong and Liu, Chris and Zhong, Zhicong and Lo, Eric and Wang, Tianzheng},
  year = {2022},
  month = jul,
  journal = {Proceedings of the VLDB Endowment},
  volume = {15},
  number = {11},
  eprint = {2207.02900},
  pages = {3004--3017},
  issn = {2150-8097},
  doi = {10.14778/3551793.3551848},
}

@inproceedings{Ge_2023_CuttingLearnedIndex,
  title = {Cutting {{Learned Index}} into {{Pieces}}: {{An In-depth Inquiry}} into {{Updatable Learned Indexes}}},
  booktitle = {2023 {{IEEE}} 39th {{International Conference}} on {{Data Engineering}} ({{ICDE}})},
  author = {Ge, Jiake and Shi, Boyu and Chai, Yanfeng and Luo, Yuanhui and Guo, Yunda and He, Yinxuan and Chai, Yunpeng},
  year = {2023},
  month = apr,
  pages = {315--327},
  issn = {2375-026X},
  doi = {10.1109/ICDE55515.2023.00031},
}

@article{Lan_2023_UpdatableLearnedIndexesa,
  title = {Updatable {{Learned Indexes Meet Disk-Resident DBMS}} - {{From Evaluations}} to {{Design Choices}}},
  author = {Lan, Hai and Bao, Zhifeng and Culpepper, J. Shane and {Borovica-Gajic}, Renata},
  year = {2023},
  month = jun,
  journal = {Proceedings of the ACM on Management of Data},
  volume = {1},
  number = {2},
  pages = {1--22},
  issn = {2836-6573},
  doi = {10.1145/3589284},
}

@article{Choi_2024_CanLearnedIndexes,
  title = {Can {{Learned Indexes}} Be {{Built Efficiently}}? {{A Deep Dive}} into {{Sampling Trade-offs}}},
  author = {Choi, Minguk and Yoo, Seehwan and Choi, Jongmoo},
  year = {2024},
  month = may,
  journal = {Proceedings of the ACM on Management of Data},
  volume = {2},
  number = {3},
  pages = {1--25},
  issn = {2836-6573},
  doi = {10.1145/3654919},
}

@article{Liu_2025_HowGoodAre,
  title = {How Good Are Multi-Dimensional Learned Indexes? {{An}} Experimental Survey},
  author = {Liu, Qiyu and Li, Maocheng and Zeng, Yuxiang and Shen, Yanyan and Chen, Lei},
  year = {2025},
  month = jan,
  journal = {The VLDB Journal},
  volume = {34},
  number = {2},
  pages = {17},
  issn = {0949-877X},
  doi = {10.1007/s00778-024-00893-6},
}

@inproceedings{Wen_2022_RandomForestDensitya,
  title = {Random {{Forest Density Estimation}}},
  booktitle = {Proceedings of the 39th {{International Conference}} on {{Machine Learning}}},
  author = {Wen, Hongwei and Hang, Hanyuan},
  year = {2022},
  month = jun,
  pages = {23701--23722},
  publisher = {PMLR},
  issn = {2640-3498},
}

@article{Liu_2018_AddingSpatialDistribution,
  title = {Adding Spatial Distribution Clue to Aggregated Vector in Image Retrieval},
  author = {Liu, Pingping and Miao, Zhuang and Guo, Huili and Wang, Yeran and Ai, Ni},
  year = {2018},
  month = feb,
  journal = {EURASIP Journal on Image and Video Processing},
  volume = {2018},
  number = {1},
  pages = {9},
  issn = {1687-5281},
  doi = {10.1186/s13640-018-0247-0},
}

@article{Liu_2018_FastIdentificationUrban,
  title = {Fast {{Identification}} of {{Urban Sprawl Based}} on {{K-Means Clustering}} with {{Population Density}} and {{Local Spatial Entropy}}},
  author = {Liu, Lingbo and Peng, Zhenghong and Wu, Hao and Jiao, Hongzan and Yu, Yang and Zhao, Jie},
  year = {2018},
  month = jul,
  journal = {Sustainability},
  volume = {10},
  number = {8},
  pages = {2683},
  issn = {2071-1050},
  doi = {10.3390/su10082683},
}

@article{Zhang_2022_CARMICacheawareLearneda,
  title = {{{CARMI}}: A Cache-Aware Learned Index with a Cost-Based Construction Algorithm},
  author = {Zhang, Jiaoyi and Gao, Yihan},
  year = {2022},
  month = jul,
  journal = {Proceedings of the VLDB Endowment},
  volume = {15},
  number = {11},
  pages = {2679--2691},
  issn = {2150-8097},
  doi = {10.14778/3551793.3551823},
}

@misc{Neufang_2024_SurrogateBasedOptimizationTechniques,
  title = {Surrogate-{{Based Optimization Techniques}} for {{Process Systems Engineering}}},
  author = {Neufang, Mathias and Pajak, Emma and van de Berg, Damien and Lee, Ye Seol and Chanona, Ehecatl Antonio del Rio},
  year = {2024},
  month = dec,
  number = {arXiv:2412.13948},
  eprint = {2412.13948},
  publisher = {arXiv},
  doi = {10.48550/arXiv.2412.13948},
}

@article{Queipo_2005_SurrogatebasedAnalysisOptimization,
  title = {Surrogate-Based Analysis and Optimization},
  author = {Queipo, Nestor V. and Haftka, Raphael T. and Shyy, Wei and Goel, Tushar and Vaidyanathan, Rajkumar and Kevin Tucker, P.},
  year = {2005},
  month = jan,
  journal = {Progress in Aerospace Sciences},
  volume = {41},
  number = {1},
  pages = {1--28},
  issn = {03760421},
  doi = {10.1016/j.paerosci.2005.02.001},
}

@misc{OSM,
   author = {{OpenStreetMap contributors}},
   title = {{OpenStreetMap Planet Dump}},
   howpublished = "\url{https://www.openstreetmap.org}",
   year = {2025},
 }

@misc{TPCH,
  author = {{TPC, Transaction Processing Performance Council}},
  title = {TPC-H: Decision Support Benchmark},
  howpublished = "\url{https://www.tpc.org/tpch/}",
  year = {2025}
}

\end{document}